# Visually Lossless Coding in HEVC: A High Bit Depth and 4:4:4 Capable JND-Based Perceptual Quantisation Technique for HEVC

Lee Prangnell

Department of Computer Science, University of Warwick, England, UK

**Abstract** — Due to the increasing prevalence of high bit depth and YCbCr 4:4:4 video data, it is desirable to develop a JND-based visually lossless coding technique which can account for high bit depth 4:4:4 data in addition to standard 8-bit precision chroma subsampled data. In this paper, we propose a Coding Block (CB)-level JND-based luma and chroma perceptual quantisation technique for HEVC named Pixel-PAQ. Pixel-PAQ exploits both luminance masking and chrominance masking to achieve JND-based visually lossless coding; the proposed method is compatible with high bit depth YCbCr 4:4:4 video data of any resolution. When applied to YCbCr 4:4:4 high bit depth video data, Pixel-PAQ can achieve vast bitrate reductions — of up to 75% (68.6% over four QP data points) — compared with a state-of-the-art luma-based JND method for HEVC named IDSQ. Moreover, the participants in the subjective evaluations confirm that visually lossless coding is successfully achieved by Pixel-PAQ (at a PSNR value of 28.04 dB in one test).

## 1.0 Introduction

Just Noticeable Distortion (JND)-based visually lossless coding is presently of considerable interest in video coding and image coding research; for example, visually lossless compression is a core consideration in the emerging JPEG-XS still image coding standard. Focusing on video compression in the HEVC standard, JND-based video coding can profoundly reduce the perceptual redundancies that are present in raw YCbCr video data. Therefore, the number of bits required to store each pixel can be considerably reduced without incurring a decrease in the perceptual quality of the reconstructed video data. As such, burdens related to data storage, transmission and bandwidth can be reduced to an extremely high degree. JND is generally defined as the maximum visibility threshold before lossy compression distortions are perceptually discernible to the Human Visual System (HVS) [1, 2]; JND has its roots in the Weber–Fechner law [3]. Even without considering JND, it is well known that raw YCbCr video data, for example, contains a high level of perceptually redundant information. To this end, the HEVC standard [4, 5] includes a multitude of advanced video coding algorithms to achieve high efficiency spatiotemporal compression of raw video data. In the lossy video coding pipeline, spatial image coding (intra-frame coding) and also Group Of Pictures (GOP)-based spatiotemporal video coding (inter-frame coding) are initially employed to dramatically reduce spatiotemporal redundancies typically inherent in all raw video sequences. Intra prediction errors [6] and inter prediction errors [7] produce luma and chroma residual values [8]. The residual values are subsequently transformed into the frequency domain by integer approximations of the Discrete Cosine Transform (DCT) and the Discrete Sine Transform (DST) [9]. The transformed residual values are then quantised using a combination of Rate Distortion Optimised Quantisation (RDOQ) and Uniform Reconstruction Quantisation (URQ) [10]. The DC transform coefficient and the low frequency and medium frequency AC transform coefficients contain the energy which is deemed as the most important in terms of reconstruction quality. Therefore, quantisation is designed to discard the least perceptually important AC coefficients (i.e., the high frequency, or low energy, AC coefficients); the degree to which high frequency AC coefficients are zeroed out is contingent upon the Quantisation Step Size (QStep). Lossless entropy coding of the quantised transform coefficients is performed by the Context Adaptive Binary Arithmetic Coding (CABAC) method; this is the stage at which the actual data compression takes place [11]. If high levels of quantisation are applied, this gives rise to a decrease in non-zero quantised coefficients, which means that the CABAC entropy coder can compress the quantised coefficients more efficiently; that is, the compressed bitstream after entropy coding will contain fewer bits.

With a focused concentration on lossy video coding in the JCT-VC HEVC HM reference codec [12], the video coding algorithms in HEVC HM are based primarily on rate-distortion theory. Consequently, visual quality measurements in HEVC lossy video coding applications are founded upon the Mean Squared Error (MSE) [13]; that is, the MSE of the reconstructed pixel data compared with the raw pixel data. It is a well established fact that the Peak Signal-to-Noise Ratio (PSNR) — which is a logarithmic visual quality metric based on MSE — has a very poor correlation with human visual perception. This is primarily due to the fact that MSE is categorised as a simple statistical risk function; it is often employed in the field of statistics for calculating the average of the squares of the deviations [13]. Therefore, it is considered to be an overly simplistic measuring tool for computing the perceptual quality of compressed video data.

In addition to the primary objective of improving coding efficiency, most lossy video coding algorithms employed in HEVC HM are designed with an emphasis on increasing the PSNR values in the compressed video data. These algorithms include Rate Distortion Optimisation (RDO) [14], RDOQ [15], Deblocking Filter (DF) [16] and Sample Adaptive Offset (SAO) [17]. Note that RDO, RDOQ, DF and SAO are effective methods in terms of increasing PSNR values for the reconstructed video; however, the PSNR-based mathematical reconstruction quality improvement attained by these techniques is perceptually negligible in terms of how the human observer interprets the perceived quality of the compressed video data. For instance, several studies have shown that a compressed video with a PSNR measure of 40 Decibels (dB), or above, typically constitutes visually lossless coding. That is, a coded video with a PSNR $\geq$ 40 dB is perceptually indistinguishable from the raw video data. Furthermore, using the example of PSNR $\geq$ 40 dB for visually lossless coding, this also implies that targeting a reconstruction quality of PSNR > 40 dB (e.g., PSNR = 50 dB) is superfluous; i.e., unnecessary bits would be wasted by achieving the superior mathematical reconstruction quality required for the PSNR = 50 dB measurement.

The key difference between JND-based video coding and video coding based on rate-distortion theory is as follows: JND techniques prioritise, above all else, the human observer with respect to assessing the reconstruction quality of a coded video. That is, instead of focusing purely on mathematically-orientated visual quality metrics including PSNR. This is because, in the end, the human observer is the ultimate judge of the visual quality of a compressed video sequence. As such, human subjective quality evaluations are critically important in terms of assessing the reconstruction quality of video sequences coded by JND-based methods. JND techniques are primarily concerned with the following core objective: To reduce bitrates, as much as possible (i.e., reduce the number of bits required to store each pixel), without incurring a perceptually discernible decrease in visual quality in the compressed video data. Note that with JND and visually lossless coding, PSNR measurements are not considered to be important in terms of quantifying the perceptual quality of a reconstructed sequence. In such cases, the PSNR metric is utilised for quantifying the degree to which PSNR values can be decreased before the associated compression-induced distortions in the coded video are perceptually discernible.

The vast majority of JND techniques in video compression applications target the spatiotemporal domain, the frequency domain or a combination of the two. Mannos' and Sakrison's pioneering work in [18] formed a useful foundation for all frequency domain luminance Contrast Sensitivity Function (CSF)-based JND techniques which target HVS-based redundancies in luminance image data. Chou's and Chen's pioneering pixel-wise JND method in [19, 20] formed the basis for several spatiotemporal domain JND contributions. The primary means by which Chou and Chen achieved pixel-wise JND are luminance-based spatial masking, contrast masking and temporal masking.

## 1.1 Overview of Related Work

In [21], Ahuma and Peterson devise the first DCT-based JND technique, in which a luminance spatial CSF is incorporated. In [22], Watson expands on Ahuma's and Peterson's work by incorporating luminance masking and contrast masking into the spatial CSF (in the frequency domain); note that power functions corresponding to Weber's law are utilised in this method. Chou and Chen develop a pioneering pixel-wise JND profile in [19], in which luminance masking and contrast masking functions are proposed for utilisation in the spatial domain (8-bit precision luma component); this method is based on average background luminance and also luminance adaptation. The authors further expand on this method in [20] by adding a temporal masking component, in which inter-frame luminance is exploited. Yang et al. in [23] propose a pixel-wise JND contribution to eradicate the overlapping effect between luminance masking and contrast masking effects. This technique also includes a filter for motion-compensated residuals, in which they employ a modified version of Chou's and Chen's spatiotemporal domain JND methods. In [24], Jia et al. present a DCT-based JND technique founded upon on a CSF-related temporal masking effect. Wei and Ngan in [25] introduce a novel DCT-based JND method for video coding, in which the authors incorporate luminance masking, contrast masking and temporal masking effects into the technique. The luminance masking component is modelled as a piecewise linear function. The contrast masking aspect is contextualised as edge and texture masking; the temporal masking component quantifies temporal frequency by taking into account motion direction. Chen and Guillemot in [26] propose a spatial domain foveated masking JND technique, which is the first time that image fixation points are taken into account in JND modelling. Moreover, this method also incorporates the luminance masking, contrast masking and temporal masking functions from Chou's and Chen's methods in [19, 20]. In [27], Naccari and Mrak propose a JND-based perceptual quantisation method (named IDSQ) which exploits luminance CSF-related spatial masking. IDSQ exploits the decreased perceptual sensitivity of the HVS to quantisation-induced compression artifacts in areas within YCbCr video data that contain high and low luma sample intensities. Y. Zhang et al. in [28] expand on Naccari's and Mrak's IDSQ technique by applying it to High Dynamic Range (HDR)-related tone-mapping applications.

As is evident in the overwhelming vast majority of JND contributions that have been previously proposed, the JND of chrominance data is typically neglected. Several JND methods reviewed in the previous paragraph share one or more of the same features including luminance masking, luminance-based contrast masking, luminance-based temporal masking and luminance-based spatial CSF. As such, if the corresponding JND techniques were to be applied to contemporary video coding applications, the JND threshold for chrominance data would be treated as identical to the JND threshold for luminance data. This is a major drawback because chrominance data is considerably different from luminance data; therefore, this leaves room for improvement. It is important and desirable to develop a comprehensive JND method that accounts for both luminance and chrominance data.

In addition to the absence of accounting for chrominance JND, other issues exist that are not considered in contemporary JND techniques. For example, the method proposed by Yang et al. and also the technique proposed by Chen and Guillemot both employ the luminance masking and contrast masking functions derived by Chou's and Chen's techniques in [19, 20]. The issue here is as follows: the psychophysical experiments undertaken by Chou and Chen were conducted in 1995-96 on obsolete visual display technologies (i.e., an SD and low resolution 19 inch CRT monitor). Therefore, Chou's and Chen's corresponding luminance masking and contrast masking functions may require revisions. This is because the derived JND visibility thresholds may prove to be significantly different if the corresponding subjective evaluations were to be performed on contemporary visual display technologies (e.g., a state-of-the-art TV or monitor which supports HD, Ultra HD, HDR, WCG and 4:4:4 video data).

Another potential issue with previously proposed JND methods — with the exception of Y. Zhang's HDR-related tone-mapping extension [28] of Naccari's and Mrak's JND-based IDSQ technique — is the fact that they are designed for raw 24-bit YCbCr data (i.e., 8-bits per channel data). This equates to the fact that most of the aforementioned empirical parameters in the luminance masking, contrast masking and temporal masking functions are designed to work with 8-bit precision data only. This may prove to be a significant issue because high bit depth data (i.e., up to 16-bits per channel data) is becoming increasingly popular in current video and image applications. Due to the increasing utilisation of 4:4:4 video data which comprise high bit depth, HD and Ultra HD characteristics, the perceptual video coding of all colour channels in such data is desirable. As per the literature review, there is presently a significant research gap. There is an absence of a JND technique which accounts for: i) Both the luminance channel and the chrominance channels; ii) The bit depth of raw video data — e.g., 8-bit, 10-bit, 12-bit and 16-bit YCbCr data; and iii) Evaluations on full chroma sampling video data (i.e., YCbCr 4:4:4) in addition to chroma subsampled data (i.e., YCbCr 4:2:0 and 4:2:2).

In this paper, a CB-level JND-based luma and chroma perceptual quantisation technique (named Pixel-PAQ) is proposed for HEVC. Pixel-PAQ is designed to perceptually increase the Y QP, the Cb QP and the Cr QP at the CB level in HEVC; this approach facilitates the JND-based perceptual coding of both luma and chroma data. One significant feature of Pixel-PAQ is that it extends Naccari's and Mrak's JND-based IDSQ technique in [27]; that is, the JND for chrominance data is accounted for in Pixel-PAQ (as opposed to luminance data only, which is the case with IDSQ). Accordingly, the proposed technique exploits both luminance masking and chrominance masking based on spatial CSF-related luminance adaptation and chrominance adaptation. In relation to the perceptual coding of chroma Cb and Cr data, Pixel-PAQ has the potential to considerably outperform Naccari's and Mrak's JND-based IDSQ technique in terms of bitrate reductions. According to the aforementioned chrominance CSF-related functions, Pixel-PAQ is designed to apply coarser levels of quantisation to Cb and Cr data when coding YCbCr 4:4:4 data and also chroma subsampled (4:2:0 and 4:2:2) data. The proposed method is particularly effective when applied to high bit depth YCbCr 4:4:4 video data, primarily because the Cb and Cr channels in high bit depth 4:4:4 data typically contain a considerable amount of perceptual redundancy due to the higher variances in the chroma channels. Moreover, compression artifacts in high variance chroma data are not conspicuous to the HVS. Therefore, the Cb and Cr data in high bit depth YCbCr 4:4:4 video sequences can be compressed much more aggressively than the Y data.

The rest of this paper is organised as follows. Section 2 includes detailed technical information on the proposed Pixel-PAQ method. Section 3 includes the evaluation, results and discussion of the proposed technique. Finally, section 4 concludes the paper.

**2.0 Pixel-PAQ: JND-Based Luminance and Chrominance Perceptual Quantisation**

Pixel-PAQ extends Naccari's and Mrak's spatial CSF-related and luminance adaptation-based IDSQ JND technique in [27]. Unlike the HDR-related tone-mapping extension of this method in [28], Pixel-PAQ focuses on extending IDSQ in which we incorporate chrominance JND in addition to accounting for high bit depth luma data and high bit depth chroma data. Both luminance masking and chrominance masking piecewise functions are employed to perceptually increase quantisation levels by virtue of JND-based modifications to the luma QStep and the chroma QSteps at the CB level. A primary objective of Pixel-PAQ is to decrease the number of perceptually insignificant non-zero luma and chroma transform coefficients. This equates to the fact that, after entropy coding, the resulting coded bitstream will contain significantly fewer bits, thus reducing bitrate and non-volatile data storage requirements. The coarser quantisation noise induced by Pixel-PAQ is indiscernible to the human observer assuming that the luma and chroma JND visibility thresholds are not exceeded.

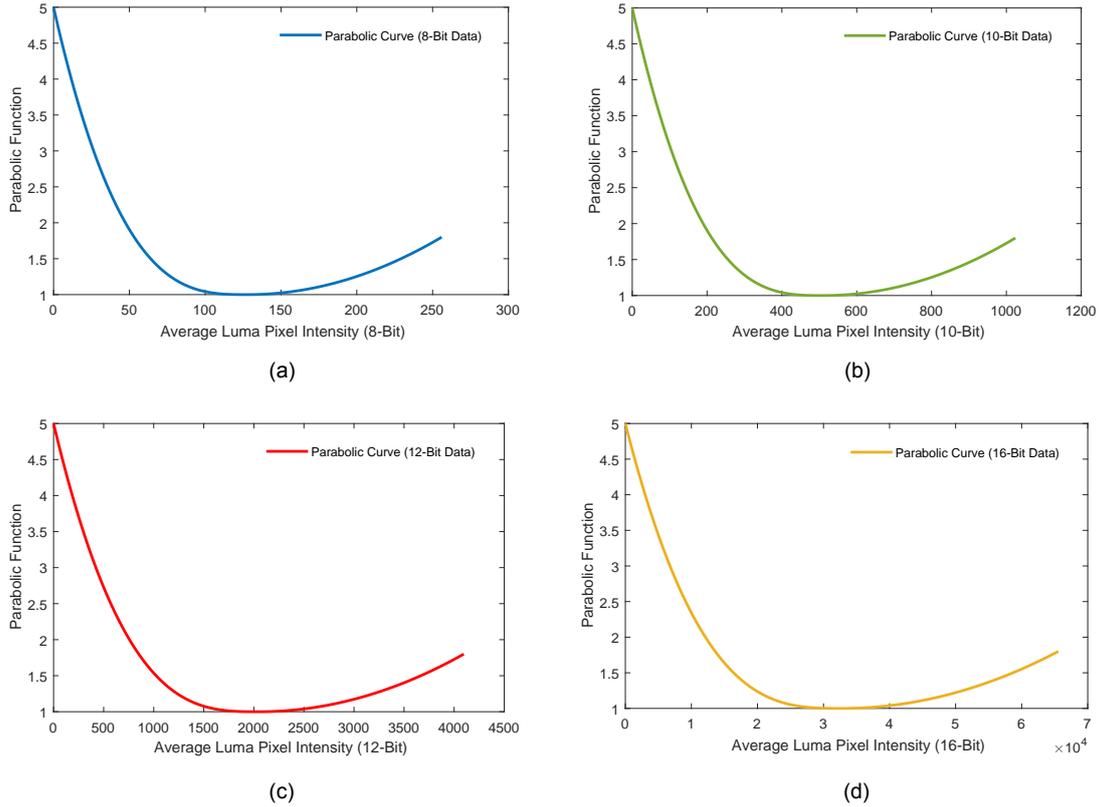

**Figure 1.** The curves derived from the parabolic function $L(\mu_Y)$ in (1). Note that the subfigures are as follows: (a) corresponds to the parabolic curve when $b = 8$ (8-bit luma data), (b) $b = 10$ (10-bit luma data), (c) $b = 12$ (12-bit luma data) and $b = 16$ (16-bit luma data). Note that, regardless of the bit depth of the luma data, the integrity of the parabolic curve is preserved.

Naccari's and Mrak's JND-based IDSQ method in [27] is founded upon the DCT-based JND technique proposed by X. Zhang et al. in [29]. In [29], X. Zhang et al. conclude that there is an intrinsic relationship between luminance adaptation, background luminance and the corresponding luma data in an image. Concerning luminance adaptation, the authors of [29] assert that the contrast threshold for luminance exhibits a parabolic curve corresponding to CSF-related grey level luminance, from which a parabolic piecewise function is derived. Naccari and Mrak employ this piecewise function and recontextualise it for application in HEVC.

2.1 JND-Based Luminance Perceptual Quantisation

In Pixel-PAQ, the aforementioned parabolic piecewise function, which also constitutes the luma JND visibility threshold, denoted as $L(\mu_Y)$, is utilised as a weight to perceptually increase the luma QStep in HEVC. Function $L(\mu_Y)$ is computed in (1):

$$L(\mu_Y) = \begin{cases} a \cdot \left(1 - \dfrac{2\mu_Y}{2^b}\right)^d + 1, & \text{if } \mu_Y \leq \dfrac{2^b}{2} \\ c \cdot \left(\dfrac{2\mu_Y}{2^b} - 1\right)^f + 1, & \text{otherwise} \end{cases} \quad (1)$$

where parameters $a$, $c$, $d$ and $f$ are set to values 2, 0.8, 3 and 2, respectively. These parameter values are selected by X. Zhang et al. in [29] to determine the shape of the spatial CSF-related luminance adaptation parabolic curve (see Figure 1).

In [29], X. Zhang et al. approximate the shape of the parabola, as shown in Figure 1 (a), based on the luminance spatial CSF psychophysical experiments conducted by Ahumada and Peterson in [21]. Somewhat dissimilar to Eq. (1) in [27], we replace value 256 with $2^b$ — where $b$ denotes the bit depth of the data — in (1) to extend the dynamic range capacity. This ensures that Pixel-PAQ is compatible with luma data of any bit depth. Furthermore, the integrity of the parabolic curve, as shown in Figure 1, is preserved regardless of the value of $b$ in (1).

Assuming that value 256 in Eq. (1) in [27] is replaced with $2^b$ in (1), $L(\mu_Y)$ can therefore be utilised in perceptual quantisation techniques for luma data of any bit depth. Furthermore, it is important to note that the mean values for the full range of luma data for any bit depth — i.e., (0+256/2) for 8-bit data, (0+1024/2) for 10-bit data, (0+4096/2) for 12-bit data and (0+65536/2) for 16-bit data — equates to a perceptually identical shade of greyscale colour in the luma component.

In (1), variable $\mu_Y$ denotes the mean raw sample value in a luma CB; $\mu_Y$ is computed in (2):

$$\mu_Y = \frac{1}{2N \times 2N} \sum_{n=1}^{2N \times 2N} w_{Y_n} \tag{2}$$

where $2N \times 2N$ denotes the number of sample values in a luma CB and where variable $w_Y$ refers to the $n^{th}$ sample value in a luma CB. To reiterate, we compute $\mu_Y$ from the original, raw sample values at the luma CB level.

There is a binary logarithmic relationship between the QP and the QStep in URQ in HEVC; this is the case for both slice-level and CB-level luma and chroma quantisation. In the luminance JND aspect of Pixel-PAQ, the primary objective is to perceptually increase the luma QStep by weighing it with $L(\mu_Y)$. In URQ in HEVC, the luma QP (denoted as $QP_Y$) and the luma QStep (denoted as $QStep_Y$) are computed in (3) and (4), respectively.

$$QP_Y(QStep_Y) = \left[6 \times \log_2(QStep_Y)\right] + 4 \tag{3}$$

$$QStep_Y(QP_Y) = 2^{\frac{QP_Y - 4}{6}} \tag{4}$$

The quantisation-induced error after the reconstruction of the luma data (denoted as $q_Y$) is only perceptually discernible if it exceeds the luma JND visibility threshold $L(\mu_Y)$. Visually lossless coding is therefore achieved if $|q_Y| \leq L(\mu_Y)$.

To reiterate, the luma QStep that incurs the maximum amount of perceptually indiscernible quantisation-induced distortion is achieved by adaptively weighing $QStep_Y$ with $L(\mu_Y)$. Therefore, the CB-level JND-based perceptual luma QStep, denoted as $PStep_Y$, is quantified in (5).

$$PStep_Y = QStep_Y \cdot \left[L(\mu_Y)\right] \tag{5}$$

Accordingly, the CB-level JND-based perceptual luma QP, denoted as $PQP_Y$, is computed in (6).

$$PQP_Y(PStep_Y) = \left[6 \times \log_2(PStep_Y)\right] + 4 \tag{6}$$

## 2.2 JND-Based Chrominance Perceptual Quantisation

In [30], Naccari and Pereira propose a JND-based quantisation matrix technique for the Advanced Video Coding (AVC) standard. In this work, the authors assert that spatial CSF-related perceptual masking is similar for both luma and chroma data; this is based on the assumption that the corresponding spatial CSFs exhibit similar properties. As such, Naccari and Pereira in [30] apply the same JND threshold for luma and chroma perceptual quantisation. Although the luminance spatial CSF and the chrominance spatial CSF share somewhat similar properties [31], there are obvious differences between the two, especially in relation to the comparative sensitivity of the HVS to achromatic data and chromatic data in compressed video data [32, 33].

In Pixel-PAQ, relatively similar piecewise functions to (1) are utilised for the CB-level JND-based perceptual quantisation of chroma Cb and Cr data. The corresponding chrominance piecewise functions, denoted as $C_{Cb}(\mu_{Cb})$ and $C_{Cr}(\mu_{Cr})$, are related to chrominance CSF; i.e., based on the relationship between luminance adaptation and its impact on chroma Cb and Cr data. The mean values of $C_{Cb}(\mu_{Cb})$ and $C_{Cr}(\mu_{Cr})$ are denoted as $\mu C_{Cb}$ and $\mu C_{Cr}$, respectively, which are utilised to perceptually weigh the Cb and Cr QSteps. Functions $C_{Cb}(\mu_{Cb})$ and $C_{Cr}(\mu_{Cr})$, which constitute the chroma Cb and Cr JND visibility thresholds, are computed in (7) and (8), respectively:

$$C_{Cb}(\mu_{Cb}) = \begin{cases} \dfrac{-\mu_{Cb} \cdot (g-1)}{h+g} & \text{if } \mu_{Cb} \leq h \\ 1 & \text{if } h < \mu_{Cb} < j \\ \dfrac{(\mu_{Cb} - j) \cdot (k-1)}{(2^b - 1 - j) + 1}, & \text{otherwise} \end{cases} \quad (7)$$

$$C_{Cr}(\mu_{Cr}) = \begin{cases} \dfrac{-\mu_{Cr} \cdot (g-1)}{h+g} & \text{if } \mu_{Cr} \leq h \\ 1 & \text{if } h < \mu_{Cr} < j \\ \dfrac{(\mu_{Cr} - j) \cdot (k-1)}{(2^b - 1 - j) + 1}, & \text{otherwise} \end{cases} \quad (8)$$

where parameters $g$, $h$, $j$ and $k$ are set to values 3, 85, 90 and 3, respectively. Similar to the way in which Naccari and Mrak adopt parameter values $a$, $c$, $d$ and $f$ in [27] for IDSQ (i.e., based on the psychophysical research conducted by X. Zhang et al. in [29]), the values for parameters $g$, $h$, $j$ and $k$ in (7) and (8) are selected based on the chrominance psychophysical experiments conducted by Wang et al. in [32]. In [32], the authors conduct psychophysical experiments regarding the overlapping effect of luminance adaptation in the chrominance Cb and Cr channels, from which the values for $g$, $h$, $j$ and $k$ are derived. It is important to note that the data in the chroma Cb and Cr channels share very similar spatial properties [32, 33]; therefore, parameter values $g$, $h$, $j$ and $k$ are employed in both (7) and (8). Variables $\mu_{Cb}$ and $\mu_{Cr}$ denote the mean raw chroma sample values in chroma Cb and Cr CBs, respectively; they are computed in (9) and (10), respectively:

$$\mu_{Cb} = \frac{1}{M} \sum_{m=1}^{M} z_{Cb_m} \quad (9)$$

$$\mu_{Cr} = \frac{1}{M} \sum_{m=1}^{M} s_{Cr_m} \quad (10)$$

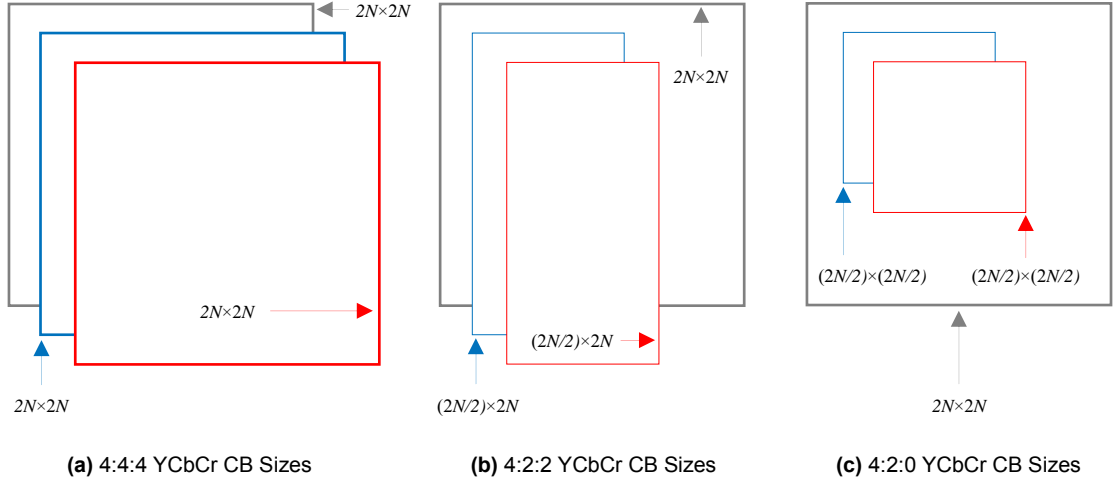

**Figure 2.** The sizes of Y, Cb and Cr CBs in a $2N \times 2N$ CU in HEVC: Y (grey), Cb (blue), Cr (red). Each subfigure specifies the size of Cb and Cr CBs for different raw video data: (a) for 4:4:4 YCbCr video data, the CB sizes for Y, Cb and Cr are all $2N \times 2N$, (b) for YCbCr 4:2:2 video data, the CB sizes are as follows: $Y_{CB} = 2N \times 2N$, $Cb_{CB} = (2N/2) \times 2N$ and $Cr_{CB} = (2N/2) \times 2N$. (c) for YCbCr 4:2:0 video data, the CB sizes are as follows: $Y_{CB} = 2N \times 2N$, $Cb_{CB} = (2N/2) \times (2N/2)$ and $Cr_{CB} = (2N/2) \times (2N/2)$.

where $M$ denotes the number of sample values in the chroma Cb and Cr CBs, variable $z_{Cb}$ refers to the $m^{th}$ sample value in a Cb CB and variable $s_{Cr}$ refers to the $m^{th}$ sample value in a Cr CB. Unlike the number of sample values in Y CBs, and also due to potential chroma subsampling, $M$ is not a fixed value. Moreover, note that Cb and Cr CBs are always identical in size regardless of the chroma sampling ratio (e.g., 4:4:4, 4:2:2 or 4:2:0) — see Figure 2. As is the case with $QP_Y$ and $QStep_Y$ in (3) and (4), respectively, in URQ there is a binary logarithmic relationship between the chroma Cb and Cr QPs (denoted as $QP_{Cb}$ and $QP_{Cr}$, respectively) and the chroma Cb and Cr QSteps (denoted as $QStep_{Cb}$ and $QStep_{Cr}$, respectively). Accordingly, $QP_{Cb}$, $QStep_{Cb}$, $QP_{Cr}$ and $QStep_{Cr}$ are computed in (11)-(14), respectively:

$$QP_{Cb}(QStep_{Cb}) = \left[6 \times \log_2(QStep_{Cb})\right] + 4 \tag{11}$$

$$QStep_{Cb}(QP_{Cb}) = 2^{\frac{QP_{Cb}-4}{6}} \tag{12}$$

$$QP_{Cr}(QStep_{Cr}) = \left[6 \times \log_2(QStep_{Cr})\right] + 4 \tag{13}$$

$$QStep_{Cr}(QP_{Cr}) = 2^{\frac{QP_{Cr}-4}{6}} \tag{14}$$

Recall that the HVS is significantly more sensitive to spatial contrast in luminance data compared with the corresponding spatial contrast sensitivity response to chromatic data. This correlates with the well established fact that the HVS is considerably less sensitive to gradations — including quantisation-induced compression artifacts — in compressed chroma data. This is the main reason why chrominance data can be quantised much more aggressively, especially high variance chroma data. To reiterate, quantisation-induced compression artifacts are vastly more perceptible in reconstructed luma data; this is primarily due to the fact that the luma channel contains all of the fine details in YCbCr pictures [34].

The quantisation-induced errors after the reconstruction of chroma Cb and Cr data, denoted as $q_{Cb}$ and $q_{Cr}$, respectively, are perceptually discernible if they exceed the chroma Cb and Cr JND visibility thresholds $C_{Cb}(\mu_{Cb})$ and $C_{Cr}(\mu_{Cr})$, respectively. Visually lossless coding is achieved if $|q_{Cb}| \leq C_{Cb}(\mu_{Cb})$ and $|q_{Cr}| \leq C_{Cr}(\mu_{Cr})$.

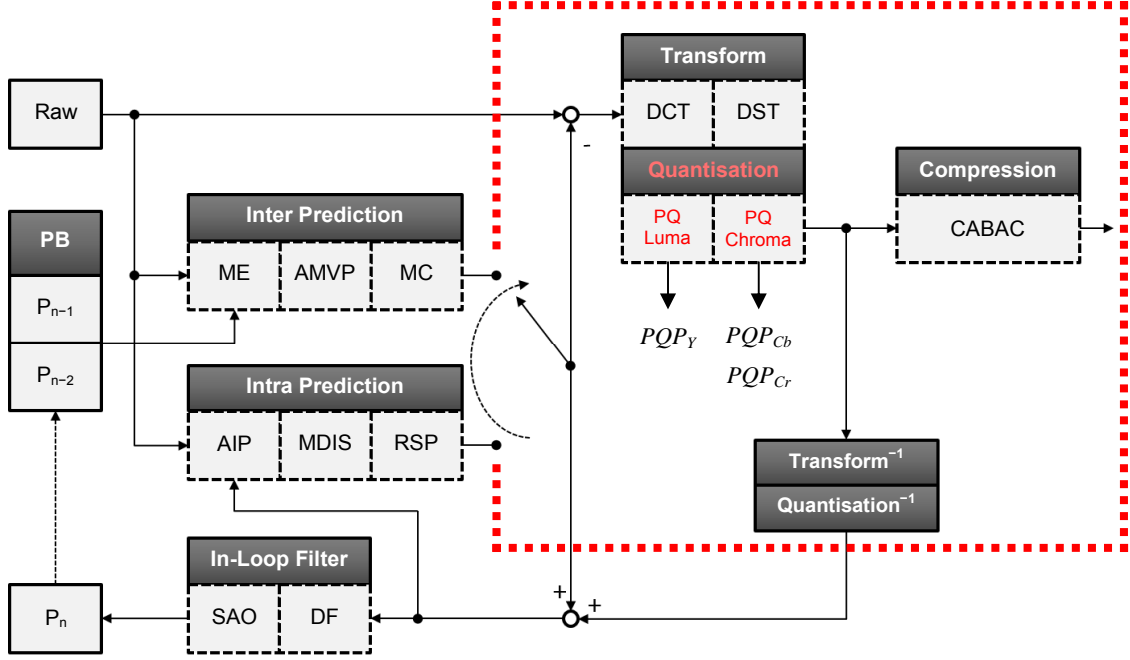

**Figure 3.** A block diagram which shows the proposed Pixel-PAQ method implemented into the JCT-VC HEVC HM encoder. The red dotted line and the red text indicate the areas within the HEVC coding pipeline in which the proposed method is implemented. Note that variables *PQP_Y*, *PQP_Cb* and *PQP_Cr* denote the perceptually adaptive QPs.

To achieve the JND-based perceptual quantisation of chroma Cb and Cr data, *QStep_Cb* and *QStep_Cr* are weighed with *μC_Cb* and *μC_Cr*, respectively. The chroma perceptual QSteps and QPs, denoted as *PStep_Cb*, *PStep_Cr*, *PQP_Cb* and *PQP_Cr*, are computed in (15)-(18), respectively.

$$PStep_{Cb} = QStep_{Cb} \cdot [\mu C_{Cb}] \qquad (15)$$

$$PQP_{Cb}(PStep_{Cb}) = [6 \times \log_2(PStep_{Cb})] + 4 \qquad (16)$$

$$PStep_{Cr} = QStep_{Cr} \cdot [\mu C_{Cr}] \qquad (17)$$

$$PQP_{Cr}(PStep_{Cr}) = [6 \times \log_2(PStep_{Cr})] + 4 \qquad (18)$$

In relation to the initial QPs utilised to evaluate Pixel-PAQ (i.e., QPs 22, 27, 32 and 37), the proposed method is implemented into HEVC HM by exploiting the CB-level chroma Cb and Cr QP offset signalling mechanism provided by JCT-VC [35, 36]. Therefore, the Cb and Cr QPs are perceptually increased at the CB level by offsetting them against *PQP_Y*. These QP and QStep offsets, denoted as *OQP_Cb*, *OStep_Cb*, *OQP_Cr* and *OStep_Cr*, respectively, are quantified in (19)-(22).

$$OQP_{Cb}(\mu C_{Cb}) = PQP_Y + [3\mu C_{Cb}] \qquad (19)$$

$$OStep_{Cb}(\mu C_{Cb}) = 2^{\frac{PQP_Y + [3\mu C_{Cb}] - 4}{6}} \qquad (20)$$

$$OQP_{Cr}(\mu C_{Cr}) = PQP_Y + [3\mu C_{Cr}] \qquad (21)$$

$$OStep_{Cr}(\mu C_{Cr}) = 2^{\frac{PQP_Y + [3\mu C_{Cr}] - 4}{6}} \qquad (22)$$

In relation to the aforementioned CB-level chroma Cb and Cr QP offset signalling technique present in the latest versions of JCT-VC HEVC HM [35, 36], this method is also exploited in our previously published perceptual quantisation contribution named FCPQ [37]. That is, we exploit the flexibility provided by JCT-VC in terms of signalling to the decoder — in the Picture Parameter Set (PPS) — chroma QP offsets at the CB level. The signalling of CB-level Cb and Cr QP offsets in the PPS proved to be particularly advantageous for FCPQ, primarily because it allows for a straightforward encoder side implementation (see Figure 3). In essence, by employing this chroma QP offset scheme, all of the CB-level quantisation-related data can be efficiently transmitted to the decoder; this ensures that the perceptually compressed video is correctly decoded and reconstructed. Furthermore, the mean raw Y, Cb and Cr sample values can be accounted for without affecting coding efficiency and computational complexity.

**3.0 Evaluation, Results and Discussion**

Pixel-PAQ is evaluated and compared with Naccari's and Mrak's JND-based IDSQ technique in [27], which has been previously proposed for the HEVC standard. It is important to affirm that IDSQ has been shown to significantly outperform both URQ and RDOQ [27] (i.e., the default scalar quantisation techniques in HEVC); furthermore, RDOQ is disabled in all tests. Pixel-PAQ is implemented into JCT-VC HEVC HM 16.7 and the method is tested on 18 official JCT-VC test sequences; namely, the proposed method is evaluated on the YCbCr 4:2:0, 4:2:2 and 4:4:4 versions of BirdsInCage, DuckAndLegs, Kimono, OldTownCross, ParkScene and Traffic. All of these sequences comprise a spatial resolution of HD 1080p (1920×1080). The 4:4:4 and 4:2:2 versions of these sequences contain a higher dynamic range (i.e., 10-bits per pixel per channel, which equates to 30-bits per pixel), whereas the 4:2:0 versions comprise 8-bits per pixel per channel. In our previously published work in [37], we provide empirical evidence that an absence of chroma subsampling in addition to a higher dynamic range for each colour channel are significantly advantageous for the perceptual quantisation of YCbCr data; this is particularly pertinent to 4:4:4 data. Therefore, this is the primary reason for employing a similar experimental setup to the one conducted in [37].

Objective visual quality evaluations are undertaken which correspond, as closely as possible, to the Common Test Conditions and Software Reference Configurations recommended by JCT-VC [38]; this is a common experimental setup utilised in contemporary HEVC research for lossy coding techniques. This includes testing techniques with four QP data points (i.e., initial QPs 22, 27, 32 and 37) with the All Intra (AI) and Random Access (RA) encoding configurations [38]. In the objective evaluation, the SSIM [39] and PNSR visual quality metrics are employed to assess the mathematical reconstruction quality of the Pixel-PAQ and IDSQ coded videos.

Due to the fact that both Pixel-PAQ and IDSQ are JND-based and HVS-orientated perceptual video coding techniques, it is of paramount importance to undertake extensive subjective visual quality evaluations in addition to the aforementioned objective visual quality evaluation. In essence, the subjective visual quality evaluations are undoubtedly the most important set of experiments in terms of measuring the perceptual quality of a compressed video sequence, especially so for visually lossless coding and JND-based techniques. As such, the United Nations' ITU-T standardised subjective evaluation procedure entitled *Subjective Video Quality Assessment Methods* (ITU-T P.910 [40]) is employed. In the ITU-T P.910 subjective evaluation, the following conditions are recommended:

- Number of participants $\geq 4$ and $\leq 40$;
- Viewing distance: $1\text{-}8 \times H$, where $H$ is the height of the TV/VDU;
- Compute Mean Opinion Score (MOS);
- Spatiotemporal analysis.

## 3.1 Bitrate Reductions and Objective Visual Quality Evaluations

**Table 1:** The overall bitrate reductions attained, per sequence, for the proposed Pixel-PAQ technique compared with IDSQ and the raw video data. In the Pixel-PAQ and IDSQ tests, the bitrates — in Kbps — are averaged over four QP data points (i.e., initial QPs 22, 27, 37 and 37). The AI results are shown on the left; the RA results are shown on the right. The green text indicates superior results (i.e., lower bitrates).

| Mean Bitrate (Kbps) – AI (YCbCr 4:2:0) | | | Mean Bitrate (Kbps) – RA (YCbCr 4:2:0) | | | |
|---|---|---|---|---|---|---|
| **Sequence** | Pixel-PAQ | IDSQ | Raw | **Sequence** | Pixel-PAQ | IDSQ | Raw |
| BirdsInCage | **18,061** | 20,286 | 1,018,880 | BirdsInCage | **1,753** | 1,942 | 1,018,880 |
| DuckAndLegs | **51,600** | 59,976 | 508,928 | DuckAndLegs | **7,859** | 8,818 | 508,928 |
| Kimono | **8,729** | 10,180 | 407,552 | Kimono | **2,422** | 2,664 | 407,552 |
| OldTownCross | **53,746** | 57,811 | 848,896 | OldTownCross | **7,625** | 7,966 | 848,896 |
| ParkScene | **25,260** | 33,778 | 407,552 | ParkScene | **3,668** | 3,892 | 407,552 |
| Traffic | **23,313** | 27,556 | 508,928 | Traffic | **3,002** | 3,292 | 508,928 |

| Mean Bitrate (Kbps) – AI (YCbCr 4:2:2) | | | Mean Bitrate (Kbps) – RA (YCbCr 4:2:2) | | | |
|---|---|---|---|---|---|---|
| **Sequence** | Pixel-PAQ | IDSQ | Raw | **Sequence** | Pixel-PAQ | IDSQ | Raw |
| BirdsInCage | **17,461** | 23,377 | 1,300,234 | BirdsInCage | **1,679** | 2,356 | 1,300,234 |
| DuckAndLegs | **53,317** | 76,647 | 635,904 | DuckAndLegs | **7,947** | 11,535 | 635,904 |
| Kimono | **8,594** | 12,478 | 509,952 | Kimono | **2,400** | 3,004 | 509,952 |
| OldTownCross | **53,389** | 67,354 | 1,090,519 | OldTownCross | **7,510** | 9,043 | 1,090,519 |
| ParkScene | **24,108** | 33,191 | 509,952 | ParkScene | **3,626** | 4,708 | 509,952 |
| Traffic | **30,861** | 36,509 | 635,904 | Traffic | **3,646** | 4,237 | 635,904 |

| Mean Bitrate (Kbps) – AI (YCbCr 4:4:4) | | | Mean Bitrate (Kbps) – RA (YCbCr 4:4:4) | | | |
|---|---|---|---|---|---|---|
| **Sequence** | Pixel-PAQ | IDSQ | Raw | **Sequence** | Pixel-PAQ | IDSQ | Raw |
| BirdsInCage | **20,278** | 40,769 | 3,911,188 | BirdsInCage | **1,830** | 5,831 | 3,911,188 |
| DuckAndLegs | **51,601** | 104,554 | 1,950,351 | DuckAndLegs | **8,390** | 22,616 | 1,950,351 |
| Kimono | **9,249** | 19,469 | 1,562,378 | Kimono | **2,495** | 4,412 | 1,562,378 |
| OldTownCross | **56,554** | 110,619 | 3,261,071 | OldTownCross | **7,764** | 17,975 | 3,261,071 |
| ParkScene | **26,048** | 44,652 | 1,562,378 | ParkScene | **3,835** | 6,703 | 1,562,378 |
| Traffic | **32,312** | 42,619 | 1,950,351 | Traffic | **3,791** | 5,171 | 1,950,351 |

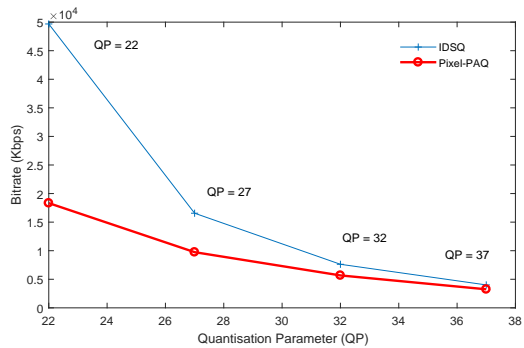
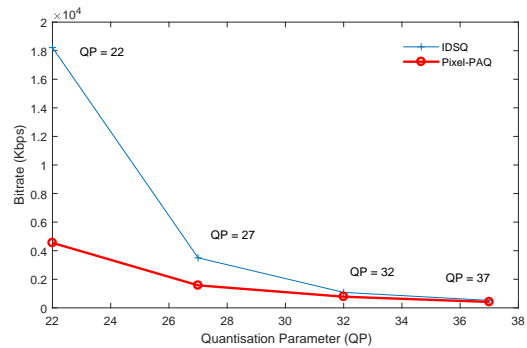

(a)    (b)

**Figure 4:** Two plots which highlight the bitrate reductions attained by Pixel-PAQ compared with IDSQ. The subfigures show the bitrate reductions achieved by IDSQ on the following sequences. Subfigure (a): Kimono 4:4:4 (AI). Subfigure (b): BirdsInCage 4:4:4 (RA).

**Table 2:** The bitrate reduction percentages (in green text), per sequence, attained for the proposed Pixel-PAQ technique compared with IDSQ. In addition, the decreased reconstruction quality (per channel) for sequences coded by Pixel-PAQ, as quantified by SSIM percentage decreases, are tabulated. The bitrate reductions are averaged over four QP data points (i.e., initial QPs 22, 27, 37 and 37). The AI results are shown on the left; the RA results are shown on the right.

**Overall Bitrate (%) Per Sequence and SSIM (%) Per Channel: Pixel-PAQ Versus IDSQ (YCbCr 4:2:0)**

| Sequence | All Intra | | | | Random Access | | | |
|---|---|---|---|---|---|---|---|---|
| | Bitrate | Y SSIM | Cb SSIM | Cr SSIM | Bitrate | Y SSIM | Cb SSIM | Cr SSIM |
| BirdsInCage | −11.0 | 0.0 | −0.4 | −0.3 | −9.7 | 0.0 | −0.3 | −0.3 |
| DuckAndLegs | −14.0 | 0.0 | −4.7 | −4.3 | −10.9 | 0.0 | −3.5 | −3.9 |
| Kimono | −14.3 | 0.0 | −0.8 | −0.6 | −9.1 | 0.0 | −0.4 | −0.4 |
| OldTownCross | −7.0 | 0.0 | −1.1 | −1.2 | −4.3 | 0.0 | −0.8 | −1.0 |
| ParkScene | −8.3 | 0.0 | −2.1 | −0.8 | −5.7 | 0.0 | −1.7 | −0.7 |
| Traffic | −11.0 | 0.0 | −1.9 | −1.1 | −8.8 | 0.0 | −1.6 | −1.0 |

**Overall Bitrate (%) Per Sequence and SSIM (%) Per Channel: Pixel-PAQ Versus IDSQ (YCbCr 4:2:2)**

| Sequence | All Intra | | | | Random Access | | | |
|---|---|---|---|---|---|---|---|---|
| | Bitrate | Y SSIM | Cb SSIM | Cr SSIM | Bitrate | Y SSIM | Cb SSIM | Cr SSIM |
| BirdsInCage | −25.3 | 0.0 | −1.0 | −0.9 | −28.7 | 0.0 | −0.4 | −0.5 |
| DuckAndLegs | −30.4 | 0.0 | −9.6 | −7.6 | −31.1 | 0.0 | −5.5 | −7.1 |
| Kimono | −31.1 | 0.0 | −1.6 | −0.8 | −20.1 | 0.0 | −0.5 | −0.5 |
| OldTownCross | −20.7 | 0.0 | −3.1 | −2.8 | −16.9 | 0.0 | −1.1 | −1.4 |
| ParkScene | −27.4 | 0.1 | −4.7 | −2.9 | −23.0 | 0.0 | −3.0 | −2.4 |
| Traffic | −15.5 | 0.0 | −3.4 | −1.8 | −13.9 | 0.0 | −2.8 | −1.5 |

**Overall Bitrate (%) Per Sequence and SSIM (%) Per Channel: Pixel-PAQ Versus IDSQ (YCbCr 4:4:4)**

| Sequence | All Intra | | | | Random Access | | | |
|---|---|---|---|---|---|---|---|---|
| | Bitrate | Y SSIM | Cb SSIM | Cr SSIM | Bitrate | Y SSIM | Cb SSIM | Cr SSIM |
| BirdsInCage | −50.3 | 0.0 | −2.8 | −0.7 | −68.6 | 0.0 | −0.4 | −0.5 |
| DuckAndLegs | −50.6 | 0.0 | −16.2 | −10.1 | −62.9 | 0.1 | −10.1 | −8.7 |
| Kimono | −52.5 | 0.0 | −3.0 | −1.0 | −43.4 | 0.0 | −0.5 | −0.6 |
| OldTownCross | −48.9 | 0.0 | −8.8 | −4.2 | −56.8 | 0.0 | −1.1 | −1.7 |
| ParkScene | −41.7 | 0.0 | −6.3 | −3.8 | −42.8 | 0.0 | −3.2 | −3.1 |
| Traffic | −24.2 | 0.0 | −3.3 | −1.4 | −26.7 | 0.0 | −3.0 | −1.8 |

In this section, the bitrate reduction results and also the mathematical reconstruction quality results are addressed. In the next sub-section the subjective evaluation results are analysed. As shown in the plots in Figure 4 and also in Table 1, Pixel-PAQ achieves exceptional bitrate reduction results on YCbCr 4:4:4 10-bit sequences in comparison with IDSQ. The most outstanding result is achieved on the BirdsInCage 4:4:4 sequence for the initial QP = 22 test using the RA encoding configuration (see Table 1 and Figure 4). In this particular test and compared with IDSQ, over 75% bitrate reductions are achieved by Pixel-PAQ; this averages out at 68.6% bitrate reductions over four initial QP values (i.e., QPs 22, 27, 32 and 37). In the RA QP = 22 test, the following bitrate reductions are attained: 4,394.99 Kbps (Pixel-PAQ) versus 17,760.41 Kbps (IDSQ) for 600 frames. In terms of data storage requirements on a non-volatile medium, the corresponding final file sizes of the compressed bitstreams are as follows: 5,368 KB (Pixel-PAQ) versus 21,683 KB (IDSQ). Furthermore, the raw BirdsInCage 4:4:4 sequence is 6.95 GB in size and the HEVC mathematically lossless coded version is 2 GB in size; the corresponding Pixel-PAQ coded bitstream is 5.24 MB in size for the RA QP = 22 test.

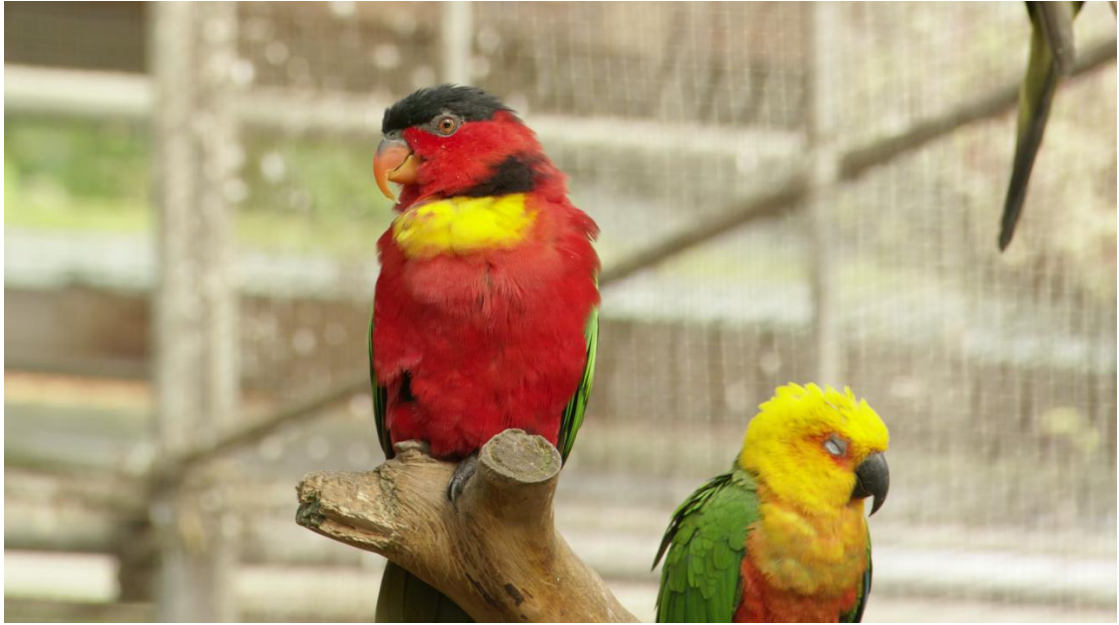

(a) Pixel-PAQ Coded Inter-Frame (RA, QP = 22)

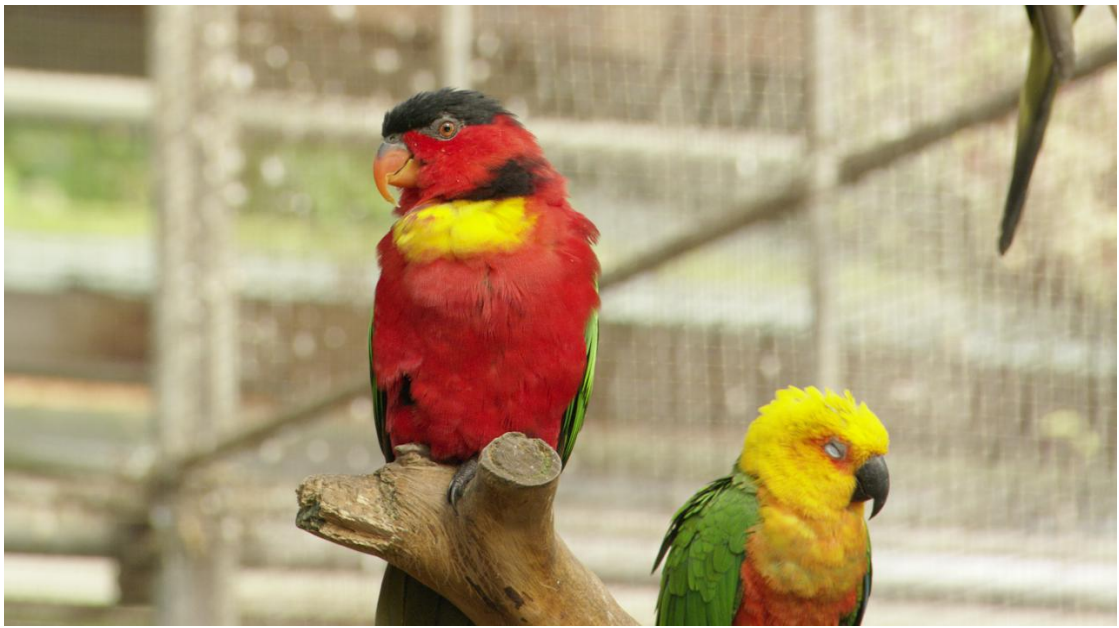

(b) Raw Data

**Figure 5:** A frame from the BirdsInCage 4:4:4 sequence. Subfigure (a) is a Pixel-PAQ coded inter-frame from this sequence (RA QP = 22 test). Subfigure (b) is the corresponding raw data. In spite of the extremely high bitrate reduction of 75% for this particular test (see Table 1 and Table 2), the Pixel-PAQ coded sequence in (a) is perceptually indistinguishable from the raw data in (b); this is also confirmed in the subjective evaluations.

In the RA QP = 22 test on this sequence, visually lossless coding is achieved by Pixel-PAQ. That is, based on the Mean Opinion Score (MOS), all four individuals who participated in the subjective evaluations could not discern any perceptible differences between the Pixel-PAQ coded version of the BirdsInCage 4:4:4 sequence and the corresponding raw sequence (the subjective evaluation results are included in sub-section 3.2). This equates to 5.24 MB (Pixel-PAQ) versus 6.95 GB (raw) for identical perceptual quality; compare Figure 5 (a) with Figure 5 (b).

**Table 3:** The bitrate reduction percentages (in green text), per sequence, attained for the proposed Pixel-PAQ technique compared with IDSQ. In addition, the decreased reconstruction quality (per channel) for sequences coded by Pixel-PAQ, as quantified by PSNR percentage decreases, are tabulated. The bitrate reductions are averaged over four QP data points (i.e., initial QPs 22, 27, 37 and 37). The AI results are shown on the left; the RA results are shown on the right.

**Overall Bitrate (%) Per Sequence and PSNR (%) Per Channel: Pixel-PAQ Versus IDSQ (YCbCr 4:2:0)**

| Sequence | All Intra | | | | Random Access | | | |
|---|---|---|---|---|---|---|---|---|
| | **Bitrate** | **Y PSNR** | **Cb PSNR** | **Cr PSNR** | **Bitrate** | **Y PSNR** | **Cb PSNR** | **Cr PSNR** |
| BirdsInCage | **−11.0** | 0.0 | −4.0 | −5.2 | **−9.7** | 0.0 | −3.2 | −4.7 |
| DuckAndLegs | **−14.0** | 0.0 | −5.7 | −5.8 | **−10.9** | 0.0 | −3.7 | −4.5 |
| Kimono | **−14.3** | 0.0 | −4.5 | −4.6 | **−9.1** | 0.0 | −3.2 | −3.0 |
| OldTownCross | **−7.0** | 0.0 | −3.5 | −4.7 | **−4.3** | 0.0 | −2.7 | −3.9 |
| ParkScene | **−8.3** | 0.0 | −6.1 | −4.4 | **−5.7** | 0.0 | −5.1 | −3.7 |
| Traffic | **−11.0** | 0.1 | −5.8 | −5.5 | **−8.8** | −0.1 | −4.7 | −4.9 |

**Overall Bitrate (%) Per Sequence and PSNR (%) Per Channel: Pixel-PAQ Versus IDSQ (YCbCr 4:2:2)**

| Sequence | All Intra | | | | Random Access | | | |
|---|---|---|---|---|---|---|---|---|
| | **Bitrate** | **Y PSNR** | **Cb PSNR** | **Cr PSNR** | **Bitrate** | **Y PSNR** | **Cb PSNR** | **Cr PSNR** |
| BirdsInCage | **−25.3** | 0.0 | −5.0 | −7.0 | **−28.7** | 0.0 | −3.4 | −6.3 |
| DuckAndLegs | **−30.4** | 0.0 | −7.5 | −7.4 | **−31.1** | 0.0 | −3.8 | −5.7 |
| Kimono | **−31.1** | 0.0 | −5.0 | −5.6 | **−20.1** | −0.1 | −3.4 | −4.3 |
| OldTownCross | **−20.7** | 0.1 | −4.9 | −5.7 | **−16.9** | 0.0 | −2.8 | −4.4 |
| ParkScene | **−27.4** | 0.0 | −6.8 | −6.1 | **−23.0** | 0.0 | −4.7 | −4.8 |
| Traffic | **−15.5** | 0.0 | −9.2 | −8.4 | **−13.9** | 0.0 | −7.4 | −7.1 |

**Overall Bitrate (%) Per Sequence and PSNR (%) Per Channel: Pixel-PAQ Versus IDSQ (YCbCr 4:4:4)**

| Sequence | All Intra | | | | Random Access | | | |
|---|---|---|---|---|---|---|---|---|
| | **Bitrate** | **Y PSNR** | **Cb PSNR** | **Cr PSNR** | **Bitrate** | **Y PSNR** | **Cb PSNR** | **Cr PSNR** |
| BirdsInCage | **−50.3** | 0.0 | −4.3 | −4.4 | **−68.6** | 0.0 | −2.1 | −3.8 |
| DuckAndLegs | **−50.6** | −0.1 | −8.9 | −7.9 | **−62.9** | 0.0 | −3.5 | −5.0 |
| Kimono | **−52.5** | 0.0 | −4.3 | −4.3 | **−43.4** | 0.0 | −2.0 | −3.3 |
| OldTownCross | **−48.9** | 0.1 | −5.5 | −4.6 | **−56.8** | 0.1 | −2.1 | −2.8 |
| ParkScene | **−41.7** | 0.1 | −6.4 | −6.0 | **−42.8** | 0.0 | −3.6 | −4.6 |
| Traffic | **−24.2** | 0.0 | −8.3 | −8.0 | **−26.7** | 0.0 | −6.1 | −6.7 |

The mean per sequence, per channel SSIM and PSNR objective evaluation results, which are extrapolated from the Pixel-PAQ versus IDSQ tests (AI and RA QP 22, 27, 32 and 37 tests), are tabulated in Table 2 and Table 3, respectively. These results confirm that the SSIM and PSNR values of the Pixel-PAQ coded sequences are typically, and necessarily, lower as compared with those obtained for the IDSQ coded sequences; this is by virtue of the nature of the JND-based chrominance masking that is inherent in the Pixel-PAQ method. In other words, compared with IDSQ, the mathematical reconstruction quality of the data in the chroma Cb and Cr channels in the Pixel-PAQ coded sequences is significantly inferior; this is due to the JND-based chrominance masking. However, according to the subjective evaluation results, these decreases in chroma reconstruction quality proved to be imperceptible to the HVS in the vast majority of cases, especially so in the QP = 22 tests. The reconstruction of luma data is not affected because the JND visibility threshold has already been reached as a result of the computations in equation (1). The mean per sequence, per QP SSIM and PSNR values are recorded in Table 4 to Table 7, as shown in the following pages.

**Table 4:** The 'per sequence' SSIM results (AI) for 'Pixel-PAQ versus the raw data' compared with 'IDSQ versus the raw data' (initial QPs 22, 27, 32 and 37). The superior SSIM results (IDSQ) are shown in green text.

**Mean SSIM Values (Per Sequence, Per QP): Pixel-PAQ Versus IDSQ (YCbCr 4:2:0) – All Intra**

| Sequence | Pixel-PAQ | | | | IDSQ | | | |
|---|---|---|---|---|---|---|---|---|
| | QP 22 | QP 27 | QP 32 | QP 37 | QP 22 | QP 27 | QP 32 | QP 37 |
| BirdsInCage | 0.9884 | 0.9852 | 0.9821 | 0.9768 | 0.9907 | 0.9879 | 0.9850 | 0.9813 |
| DuckAndLegs | 0.9320 | 0.9005 | 0.8765 | 0.8369 | 0.9613 | 0.9195 | 0.8884 | 0.8509 |
| Kimono | 0.9227 | 0.9069 | 0.8926 | 0.8662 | 0.9396 | 0.9235 | 0.9063 | 0.8853 |
| OldTownCross | 0.9015 | 0.8534 | 0.8222 | 0.7811 | 0.9157 | 0.8660 | 0.8337 | 0.7958 |
| ParkScene | 0.9399 | 0.9138 | 0.8803 | 0.8292 | 0.9555 | 0.9302 | 0.8937 | 0.8463 |
| Traffic | 0.9461 | 0.9226 | 0.8953 | 0.8493 | 0.9615 | 0.9411 | 0.9128 | 0.8754 |

**Mean SSIM Values (Per Sequence, Per QP): Pixel-PAQ Versus IDSQ (YCbCr 4:2:2) – All Intra**

| Sequence | Pixel-PAQ | | | | IDSQ | | | |
|---|---|---|---|---|---|---|---|---|
| | QP 22 | QP 27 | QP 32 | QP 37 | QP 22 | QP 27 | QP 32 | QP 37 |
| BirdsInCage | 0.9860 | 0.9823 | 0.9763 | 0.9641 | 0.9893 | 0.9860 | 0.9828 | 0.9778 |
| DuckAndLegs | 0.9021 | 0.8668 | 0.8357 | 0.7911 | 0.9617 | 0.9098 | 0.8610 | 0.8160 |
| Kimono | 0.8990 | 0.8814 | 0.8574 | 0.8128 | 0.9283 | 0.9026 | 0.8836 | 0.8570 |
| OldTownCross | 0.8626 | 0.8166 | 0.7762 | 0.7107 | 0.9058 | 0.8362 | 0.8007 | 0.7605 |
| ParkScene | 0.8682 | 0.8383 | 0.8000 | 0.7430 | 0.9146 | 0.8676 | 0.8308 | 0.7879 |
| Traffic | 0.9506 | 0.9206 | 0.8713 | 0.7864 | 0.9675 | 0.9456 | 0.9111 | 0.8569 |

**Mean SSIM Values (Per Sequence, Per QP): Pixel-PAQ Versus IDSQ (YCbCr 4:4:4) – All Intra**

| Sequence | Pixel-PAQ | | | | IDSQ | | | |
|---|---|---|---|---|---|---|---|---|
| | QP 22 | QP 27 | QP 32 | QP 37 | QP 22 | QP 27 | QP 32 | QP 37 |
| BirdsInCage | 0.9790 | 0.9759 | 0.9717 | 0.9653 | 0.9856 | 0.9803 | 0.9761 | 0.9721 |
| DuckAndLegs | 0.8897 | 0.8396 | 0.8086 | 0.7697 | 0.9709 | 0.9183 | 0.8479 | 0.7924 |
| Kimono | 0.8792 | 0.8635 | 0.8437 | 0.8156 | 0.9279 | 0.8864 | 0.8642 | 0.8407 |
| OldTownCross | 0.8001 | 0.7515 | 0.7205 | 0.6823 | 0.9135 | 0.8122 | 0.7391 | 0.6992 |
| ParkScene | 0.8554 | 0.8268 | 0.7940 | 0.7527 | 0.9212 | 0.8623 | 0.8197 | 0.7774 |
| Traffic | 0.9535 | 0.9282 | 0.8888 | 0.8271 | 0.9686 | 0.9473 | 0.9157 | 0.8682 |

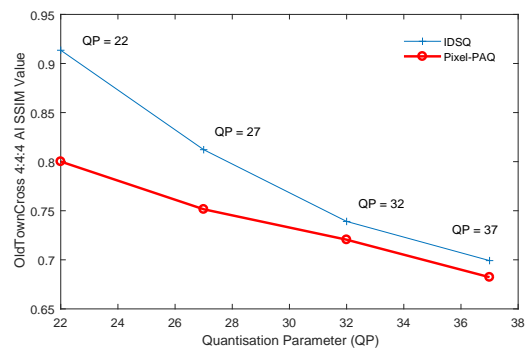 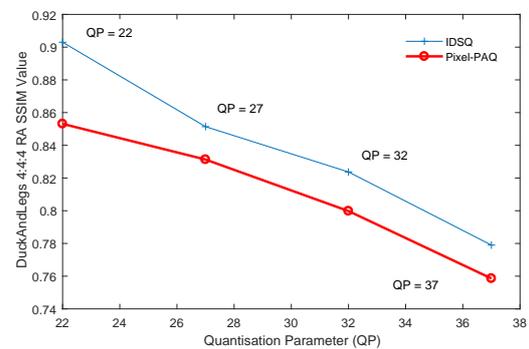

(a)      (b)

**Figure 6:** Two plots which highlight the inferior mathematical reconstruction quality of Pixel-PAQ coded sequences versus IDSQ coded sequences, over four QP data points (i.e., QPs 22, 27, 32 and 37), as quantified by the SSIM metric. Subfigure (a): OldTownCross 4:4:4 (AI). Subfigure (b): DuckAndLegs 4:4:4 (RA).

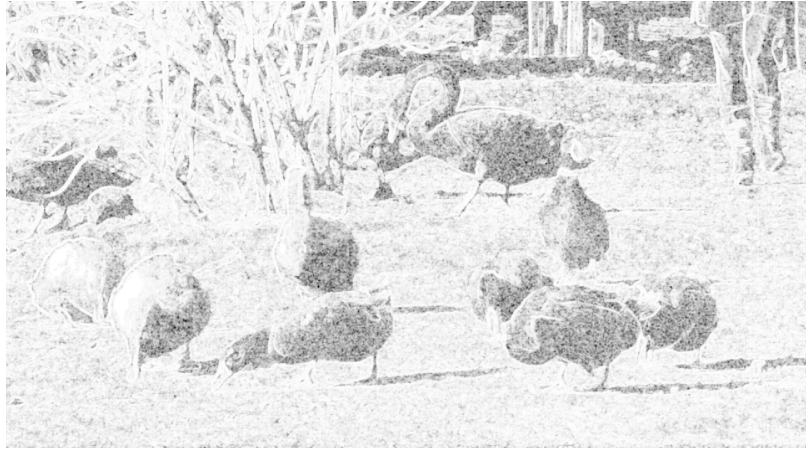

(a) Luma Channel

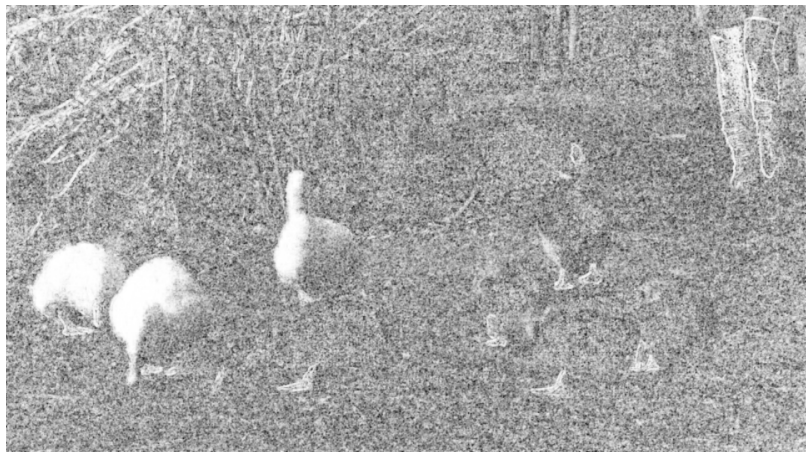

(b) Chroma Cb Channel

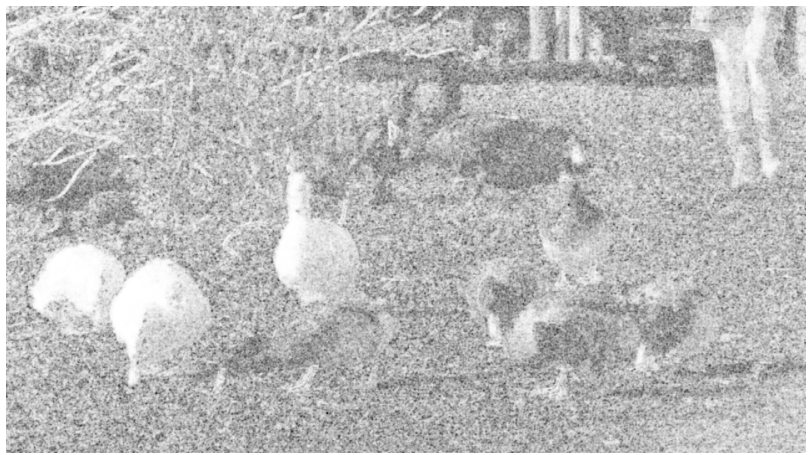

(c) Chroma Cr Channel

**Figure 7:** The SSIM Index Map (structural reconstruction errors) of a Pixel-PAQ coded intra-frame (AI QP = 22 test) versus the raw data (DuckAndLegs 4:4:4 sequence). In subfigures (a), (b) and (c), respectively, the luma (Y), chroma Cb and chroma Cr structural reconstruction errors are shown separately.

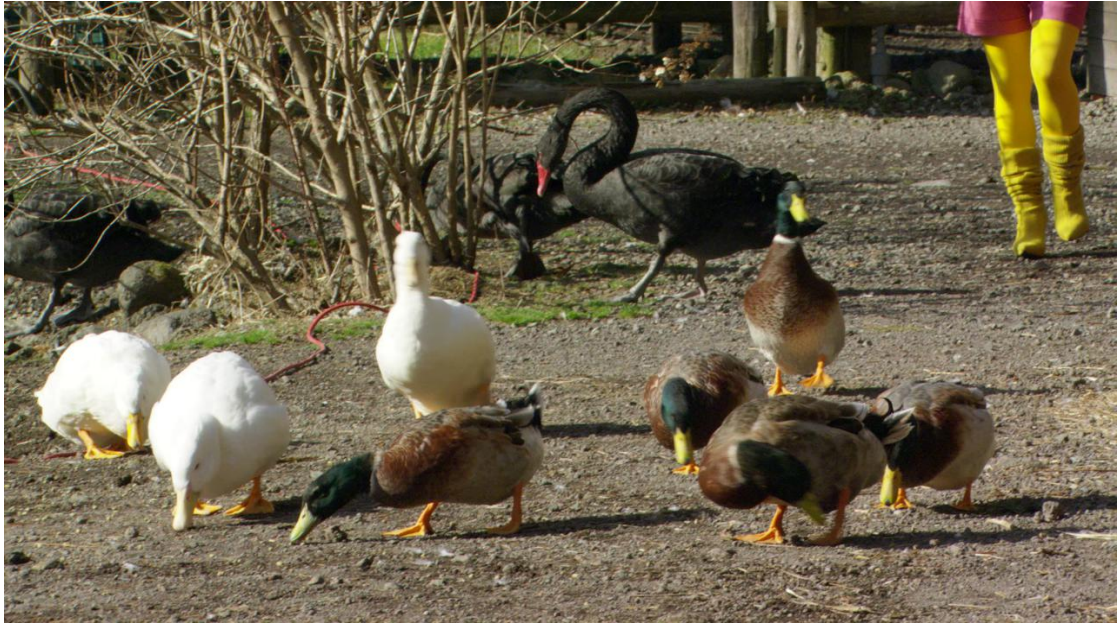

(a) Pixel-PAQ Coded Intra-Frame (AI, QP = 22)

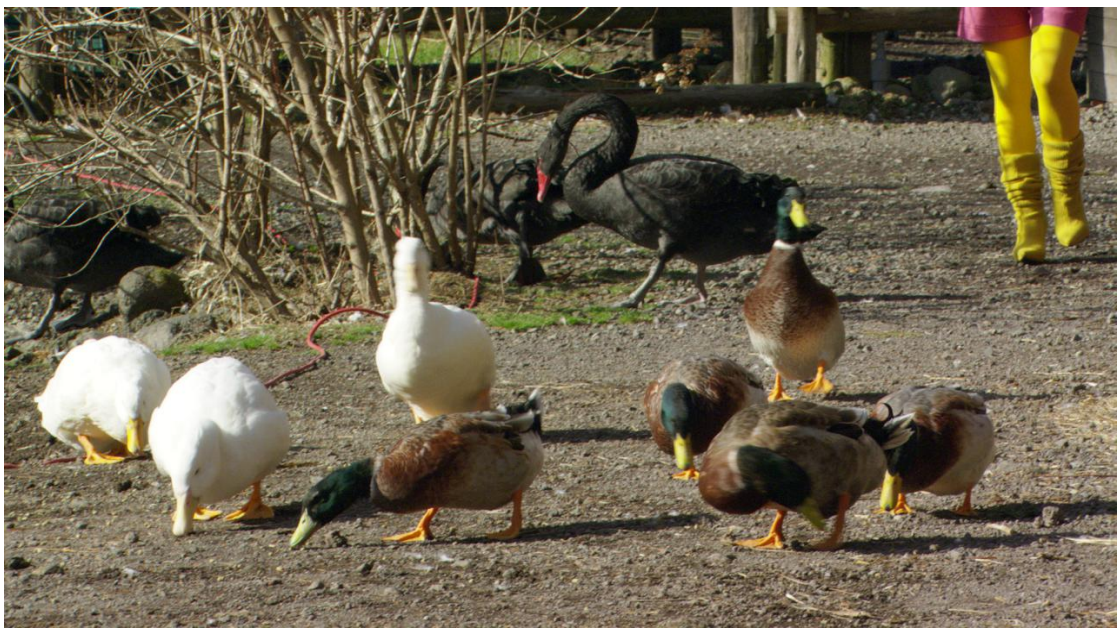

(b) Raw Data

**Figure 8:** A frame from the DuckAndLegs 4:4:4 sequence. Subfigure (a) is a Pixel-PAQ coded intra-frame from this sequence (AI QP = 22 test). Subfigure (b) is the corresponding raw data. Note that, despite the poor mathematical reconstruction quality of the data in the chroma Cb and Cr channels, as quantified by SSIM (see Figure 7), the Pixel-PAQ coded sequence in (a) is perceptually indistinguishable from the raw data in (b); this is confirmed in the subjective evaluations.

As shown in Figure 7, the structural reconstruction errors are concentrated mostly in the high variance regions in the Y, Cb and Cr channels. This is primarily because the HVS is less capable of detecting quantisation-induced compression artifacts in high spatial variance regions of compressed luma and chroma data [37]. Therefore, in spite of the reconstruction errors shown in Figure 7, visually lossless coding is attained by Pixel-PAQ in both the AI QP = 22 test — and also RA QP = 22 test on the DuckAndLegs 4:4:4 sequence. This is confirmed in the subjective evaluations; for a comparison, refer to Figure 8 (a) versus Figure 8 (b).

**Table 5:** The 'per sequence' SSIM results (RA) for 'Pixel-PAQ versus the raw data' compared with 'IDSQ versus the raw data' (initial QPs 22, 27, 32 and 37). The superior SSIM results are shown in green text.

**Mean SSIM Values (Per Sequence, Per QP): Pixel-PAQ Versus IDSQ (YCbCr 4:2:0) – Random Access**

| Sequence | Pixel-PAQ | | | | IDSQ | | | |
|---|---|---|---|---|---|---|---|---|
| | QP 22 | QP 27 | QP 32 | QP 37 | QP 22 | QP 27 | QP 32 | QP 37 |
| BirdsInCage | 0.9883 | 0.9860 | 0.9832 | 0.9782 | 0.9898 | 0.9882 | 0.9857 | 0.9825 |
| DuckAndLegs | 0.9060 | 0.8897 | 0.8666 | 0.8250 | 0.9196 | 0.9035 | 0.8770 | 0.8381 |
| Kimono | 0.9136 | 0.9002 | 0.8836 | 0.8584 | 0.9251 | 0.9120 | 0.8947 | 0.8749 |
| OldTownCross | 0.8565 | 0.8445 | 0.8275 | 0.7930 | 0.8640 | 0.8543 | 0.8374 | 0.8056 |
| ParkScene | 0.9361 | 0.9112 | 0.8780 | 0.8284 | 0.9480 | 0.9245 | 0.8909 | 0.8459 |
| Traffic | 0.9463 | 0.9267 | 0.9025 | 0.8602 | 0.9573 | 0.9408 | 0.9167 | 0.8838 |

**Mean SSIM Values (Per Sequence, Per QP): Pixel-PAQ Versus IDSQ (YCbCr 4:2:2) – Random Access**

| Sequence | Pixel-PAQ | | | | IDSQ | | | |
|---|---|---|---|---|---|---|---|---|
| | QP 22 | QP 27 | QP 32 | QP 37 | QP 22 | QP 27 | QP 32 | QP 37 |
| BirdsInCage | 0.9856 | 0.9830 | 0.9780 | 0.9673 | 0.9869 | 0.9856 | 0.9831 | 0.9789 |
| DuckAndLegs | 0.8794 | 0.8591 | 0.8292 | 0.7856 | 0.8954 | 0.8797 | 0.8512 | 0.8065 |
| Kimono | 0.8954 | 0.8766 | 0.8538 | 0.8164 | 0.9016 | 0.8907 | 0.8734 | 0.8492 |
| OldTownCross | 0.8171 | 0.8034 | 0.7798 | 0.7324 | 0.8220 | 0.8145 | 0.7977 | 0.7646 |
| ParkScene | 0.8665 | 0.8421 | 0.8079 | 0.7535 | 0.8816 | 0.8625 | 0.8322 | 0.7919 |
| Traffic | 0.9519 | 0.9276 | 0.8844 | 0.8108 | 0.9634 | 0.9460 | 0.9171 | 0.8693 |

**Mean SSIM Values (Per Sequence, Per QP): Pixel-PAQ Versus IDSQ (YCbCr 4:4:4) – Random Access**

| Sequence | Pixel-PAQ | | | | IDSQ | | | |
|---|---|---|---|---|---|---|---|---|
| | QP 22 | QP 27 | QP 32 | QP 37 | QP 22 | QP 27 | QP 32 | QP 37 |
| BirdsInCage | 0.9777 | 0.9759 | 0.9725 | 0.9666 | 0.9783 | 0.9774 | 0.9756 | 0.9724 |
| DuckAndLegs | 0.8531 | 0.8313 | 0.7998 | 0.7586 | 0.9030 | 0.8514 | 0.8237 | 0.7790 |
| Kimono | 0.8702 | 0.8574 | 0.8386 | 0.8127 | 0.8761 | 0.8673 | 0.8522 | 0.8316 |
| OldTownCross | 0.7504 | 0.7398 | 0.7205 | 0.6868 | 0.7492 | 0.7467 | 0.7326 | 0.7035 |
| ParkScene | 0.8534 | 0.8303 | 0.7994 | 0.7581 | 0.8637 | 0.8473 | 0.8204 | 0.7825 |
| Traffic | 0.9533 | 0.9322 | 0.8983 | 0.8433 | 0.9620 | 0.9458 | 0.9191 | 0.8771 |

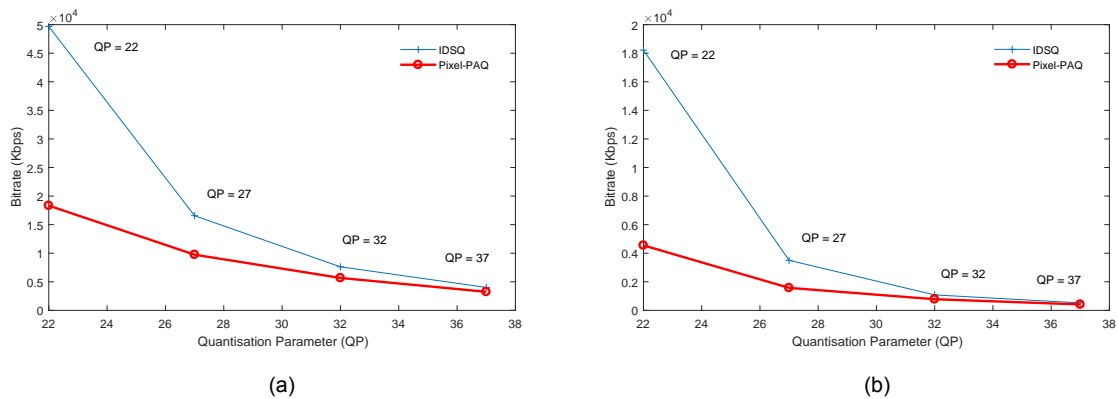

(a)      (b)

**Figure 9.** Two plots which highlight the bitrate reductions attained by Pixel-PAQ compared with IDSQ. Subfigure (a) shows the bitrate reductions achieved by Pixel-PAQ on the Kimono 4:4:4 sequence using the AI encoding configuration. Subfigure (b) shows the bitrate reductions achieved by Pixel-PAQ on the BirdsInCage 4:4:4 sequence using the RA encoding configuration.

**Table 6:** The 'per sequence' PSNR (dB) results (AI) for 'Pixel-PAQ versus the raw data' compared with 'IDSQ versus the raw data' (initial QPs 22, 27, 32 and 37). The superior SSIM results are shown in green text.

**Mean PSNR (dB) Per Sequence, Per QP: Pixel-PAQ Versus IDSQ (YCbCr 4:2:0) – All Intra**

| Sequence | Pixel-PAQ | | | | IDSQ | | | |
|---|---|---|---|---|---|---|---|---|
| | QP 22 | QP 27 | QP 32 | QP 37 | QP 22 | QP 27 | QP 32 | QP 37 |
| BirdsInCage | 37.5180 | 36.0085 | 34.8799 | 33.3267 | 38.8655 | 37.1534 | 35.7128 | 34.2626 |
| DuckAndLegs | 31.7584 | 29.8392 | 28.6015 | 26.9424 | 34.8509 | 31.3556 | 29.3250 | 27.5127 |
| Kimono | 35.8659 | 34.5481 | 33.5125 | 31.8994 | 37.5313 | 35.9396 | 34.4053 | 32.7964 |
| OldTownCross | 33.8187 | 31.8953 | 30.6139 | 29.0512 | 35.0655 | 32.7264 | 31.1386 | 29.5463 |
| ParkScene | 34.6698 | 32.5783 | 30.7900 | 28.7431 | 36.9332 | 34.0121 | 31.4805 | 29.2620 |
| Traffic | 35.2647 | 33.1950 | 31.5932 | 29.5243 | 37.2091 | 34.8116 | 32.5567 | 30.4404 |

**Mean PSNR (dB) Per Sequence, Per QP: Pixel-PAQ Versus IDSQ (YCbCr 4:2:2) – All Intra**

| Sequence | Pixel-PAQ | | | | IDSQ | | | |
|---|---|---|---|---|---|---|---|---|
| | QP 22 | QP 27 | QP 32 | QP 37 | QP 22 | QP 27 | QP 32 | QP 37 |
| BirdsInCage | 36.6285 | 35.2130 | 33.6314 | 31.6704 | 38.2948 | 36.5251 | 35.1635 | 33.6036 |
| DuckAndLegs | 30.3503 | 28.6050 | 27.2242 | 25.8192 | 34.5193 | 30.7285 | 28.4722 | 26.6346 |
| Kimono | 35.0233 | 33.7931 | 32.3663 | 30.5674 | 36.7747 | 35.2219 | 33.8264 | 32.1149 |
| OldTownCross | 32.5675 | 30.9991 | 29.6107 | 27.9068 | 34.4299 | 31.9390 | 30.5069 | 28.9463 |
| ParkScene | 32.5660 | 30.8475 | 29.2602 | 27.6884 | 35.2745 | 32.7290 | 30.6848 | 28.8019 |
| Traffic | 34.9795 | 32.4327 | 29.9572 | 27.5877 | 37.6387 | 34.6954 | 31.9075 | 29.2406 |

**Mean PSNR (dB) Per Sequence, Per QP: Pixel-PAQ Versus IDSQ (YCbCr 4:4:4) – All Intra**

| Sequence | Pixel-PAQ | | | | IDSQ | | | |
|---|---|---|---|---|---|---|---|---|
| | QP 22 | QP 27 | QP 32 | QP 37 | QP 22 | QP 27 | QP 32 | QP 37 |
| BirdsInCage | 34.8927 | 34.0048 | 33.0221 | 31.7821 | 37.2223 | 35.1969 | 33.8851 | 32.8186 |
| DuckAndLegs | 29.5754 | 27.8507 | 26.6800 | 25.4676 | 35.1345 | 30.5158 | 27.9026 | 26.1863 |
| Kimono | 34.3420 | 33.4135 | 32.2602 | 30.8592 | 36.4072 | 34.5038 | 33.3406 | 31.9085 |
| OldTownCross | 30.8036 | 29.7243 | 28.7834 | 27.6085 | 34.0635 | 30.8971 | 29.3628 | 28.1692 |
| ParkScene | 32.5565 | 30.9406 | 29.4304 | 27.9728 | 35.4066 | 32.7230 | 30.7505 | 28.8994 |
| Traffic | 35.3045 | 32.8423 | 30.4738 | 28.2057 | 37.8411 | 34.8627 | 32.1028 | 29.4800 |

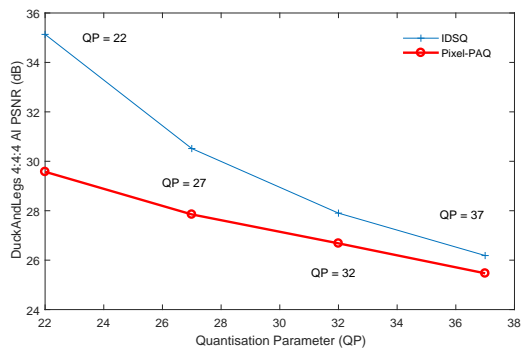 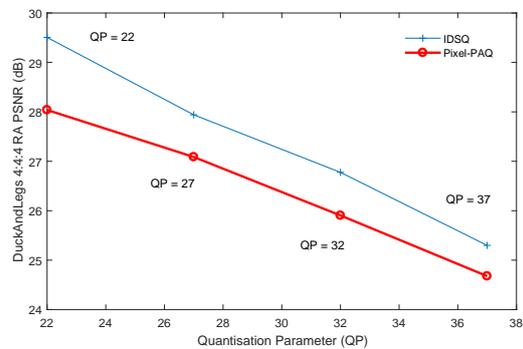

(a)          (b)

**Figure 10:** Two plots which highlight the inferior mathematical reconstruction quality of Pixel-PAQ coded sequences versus IDSQ coded sequences, over four QP data points (i.e., QPs 22, 27, 32 and 37), using the SSIM metric. Subfigure (a): DuckAndLegs 4:4:4 (AI). Subfigure (b): DuckAndLegs 4:4:4 (RA).

**Table 7:** The 'per sequence' PSNR (dB) results (RA) for 'Pixel-PAQ versus the raw data' compared with 'IDSQ versus the raw data' (initial QPs 22, 27, 32 and 37). The superior SSIM results are shown in green text.

**Mean PSNR (dB) Per Sequence, Per QP: Pixel-PAQ Versus IDSQ (YCbCr 4:2:0) – Random Access**

| Sequence | Pixel-PAQ | | | | IDSQ | | | |
|---|---|---|---|---|---|---|---|---|
| | QP 22 | QP 27 | QP 32 | QP 37 | QP 22 | QP 27 | QP 32 | QP 37 |
| BirdsInCage | 37.3938 | 36.3239 | 35.1878 | 33.6965 | 38.1668 | 37.2331 | 35.9925 | 34.6063 |
| DuckAndLegs | 30.2049 | 29.0677 | 27.8971 | 26.3244 | 31.4326 | 30.0737 | 28.4649 | 26.8022 |
| Kimono | 35.1974 | 34.0136 | 32.6811 | 31.1212 | 36.2523 | 34.8071 | 33.2019 | 31.6658 |
| OldTownCross | 32.3403 | 31.6902 | 30.8602 | 29.4741 | 32.8560 | 32.2676 | 31.3361 | 29.9690 |
| ParkScene | 34.2629 | 32.4221 | 30.7282 | 28.8326 | 35.7195 | 33.4312 | 31.3018 | 29.3340 |
| Traffic | 34.9042 | 33.2256 | 31.7335 | 29.8033 | 36.3045 | 34.3904 | 32.5393 | 30.6701 |

**Mean PSNR (dB) Per Sequence, Per QP: Pixel-PAQ Versus IDSQ (YCbCr 4:2:2) – Random Access**

| Sequence | Pixel-PAQ | | | | IDSQ | | | |
|---|---|---|---|---|---|---|---|---|
| | QP 22 | QP 27 | QP 32 | QP 37 | QP 22 | QP 27 | QP 32 | QP 37 |
| BirdsInCage | 36.4532 | 35.5006 | 34.0949 | 32.1827 | 36.9894 | 36.4115 | 35.3608 | 33.9560 |
| DuckAndLegs | 29.1533 | 27.9895 | 26.7018 | 25.3548 | 30.2635 | 29.2285 | 27.7411 | 26.0493 |
| Kimono | 34.4255 | 33.2542 | 31.8061 | 30.1685 | 35.2015 | 34.1193 | 32.6956 | 31.1283 |
| OldTownCross | 31.4985 | 30.8881 | 29.9111 | 28.5249 | 31.8896 | 31.4928 | 30.6906 | 29.3754 |
| ParkScene | 32.1890 | 30.8004 | 29.3647 | 27.8868 | 33.4481 | 32.0739 | 30.4624 | 28.7974 |
| Traffic | 34.7826 | 32.6703 | 30.4097 | 28.1447 | 36.4386 | 34.3156 | 32.0356 | 29.5884 |

**Mean PSNR (dB) Per Sequence, Per QP: Pixel-PAQ Versus IDSQ (YCbCr 4:4:4) – Random Access**

| Sequence | Pixel-PAQ | | | | IDSQ | | | |
|---|---|---|---|---|---|---|---|---|
| | QP 22 | QP 27 | QP 32 | QP 37 | QP 22 | QP 27 | QP 32 | QP 37 |
| BirdsInCage | 34.5736 | 34.1050 | 33.2660 | 32.1066 | 34.7232 | 34.4621 | 33.9179 | 33.0211 |
| DuckAndLegs | 28.0381 | 27.0870 | 25.9042 | 24.6768 | 29.5056 | 27.9400 | 26.7760 | 25.2985 |
| Kimono | 33.7630 | 32.8515 | 31.6309 | 30.2587 | 34.1753 | 33.4081 | 32.2495 | 30.9117 |
| OldTownCross | 30.0597 | 29.6798 | 29.0180 | 28.0228 | 30.1151 | 29.9754 | 29.4715 | 28.5145 |
| ParkScene | 32.1325 | 30.8363 | 29.4403 | 28.0782 | 32.9975 | 31.8978 | 30.4856 | 28.8733 |
| Traffic | 34.8858 | 32.9088 | 30.8086 | 28.6553 | 36.2828 | 34.2486 | 32.0847 | 29.8005 |

Identical in context to the SSIM results shown in Table 4 and Table 5, and as expected, the PSNR results attained by Pixel-PAQ are inferior to those achieved by IDSQ in all tests (see Table 6 and Table 7). In essence, this equates to the fact that the PSNR and SSIM objective visual quality metrics do not correspond with how the Pixel-PAQ coded sequences are subjectively perceived in realistic viewing situations, as confirmed in the subjective evaluations. To reiterate, objective visual quality metrics, including PSNR and SSIM, are employed for an entirely different purpose in perceptually-orientated, visually lossless and JND-based video compression techniques. Recall that the main objective with perceptual video coding methods is as follows: to achieve the largest bitrate reduction possible without incurring a perceptually discernible decrease of visual quality in the compressed video data. The human observer is the ultimate judge of visual quality; therefore, the objective visual quality values, regardless of the metric employed (e.g., PSNR and SSIM), are somewhat irrelevant. For instance, in the RA QP = 22 test conducted on the DuckAndLegs 4:4:4 sequence, the reconstruction quality measurement achieved — as quantified by PSNR — is approximately 28.04 dB (see Table 7). However, visually lossless coding is still achieved at this extremely low PSNR value.

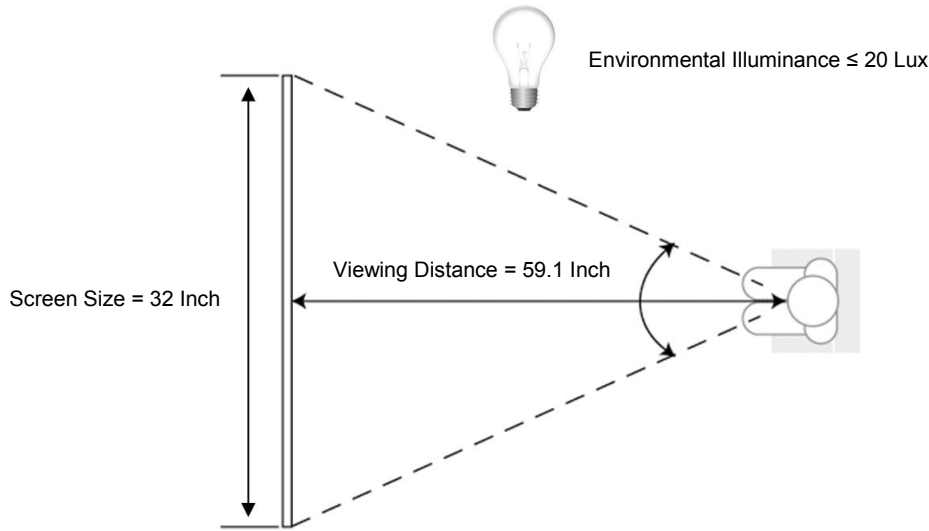

**Figure 11:** A diagram which illustrates the viewing conditions of the subjective evaluations. This includes the screen size of the TV/VDU, the viewing distance of the participant from the TV/VDU and the environmental illuminance.

3.2 Subjective Visual Quality Evaluations

As regards the subjective evaluations, the main objective is as follows: to ascertain if Pixel-PAQ coded sequences differ from IDSQ coded sequences and the raw video data in terms of subjective perceptual quality; this is quantified by the Mean Opinion Score (MOS). MOS is employed such that each participant rates the comparative reconstruction quality of Pixel-PAQ coded video data according to following MOS values: 5 = Imperceptible (Visually Lossless); 4 = Very Slightly Perceptible; 3 = Moderately Perceptible; 2 = Significantly Perceptible; 1 = Extremely Obvious.

In accordance with the conditions specified in the ITU-R Rec. P.910 [40] subjective evaluation procedures, four individuals participated in all tests; recall that ITU-R Rec. P.910 stipulates a minimum of four participants. Furthermore, the viewing distance of the participants from the TV/VDU is 1.5m in all evaluations (1.5m ≈ 59.1 inch). Therefore, the height $H$ of the TV/VDU is 15.7 inch and the viewing distance is approximately $4 \times H$. The lighting characteristics in relation to the environmental illumination were ≤ 20 lux (see Figure 11).

The four participants involved in the subjective evaluation procedure consisted of a mean age of approximately 36 years old. In terms of inclusion criteria, it is important to note that self-reported visual acuity was assessed prior to the commencement of the subjective evaluations. This is important because it ensures that the participants are able to reasonably assess the compressed video data without the need to consider visual distortions caused by ophthalmological ailments. As such, each participant involved in the assessment reported normal vision or corrected normal vision (i.e., an absence of severe ophthalmological conditions which typically affect central and peripheral vision including macular degeneration, photophobia, cataracts and glaucoma).

The computer hardware in the experimental setup consists of a desktop PC which contains an Intel Core i7-4770 CPU — 4 cores and 8 threads — running at 3.4 GHz per core. The volatile memory in the PC is as follows: 24 GB of Double Data Rate (Type 3) Synchronous Dynamic Random Access Memory (DDR3 SDRAM) running at a DRAM frequency of 680 MHz. The Graphics Processing Unit (GPU) installed in the PC is an NVIDIA GeForce 750 Ti. Note that the NVIDIA GeForce 750 Ti is an Ultra HD 4K and HDR-capable GPU that can support YCbCr 4:4:4 and RGB data of bit depths up to 12-bits per colour channel (i.e., 36-bits per pixel). Therefore, the higher dynamic ranges and also the lack of chroma subsampling in the raw 4:4:4 sequences are discernible (compared with the subsampled 4:2:0 8-bit sequences). The visual display hardware on which the subjective valuations are conducted is as follows: 1080p HD 32 Inch Samsung F5500 LED Smart TV/VDU.

**Table 8.** The MOS results, rounded to the nearest integer, of four participants in the subjective evaluations for Pixel-PAQ versus IDSQ. The sequences viewed are coded with initial QPs 22, 27, 32 and 37 on the YCbCr 4:2:0, 4:2:2 4:4:4 versions of the following sequences: BirdsInCage, DuckAndLegs Kimono, OldTownCross, ParkScene and Traffic compressed using the AI and RA configurations. The AI results are shown on the left; the RA results are shown on the right.

**Rounded Mean Opinion Score (Spatiotemporal Subjective Evaluation) – Pixel-PAQ versus IDSQ**

| Sequence | YCbCr 4:2:0 All Intra | | | | YCbCr 4:2:0 Random Access | | | |
|---|---|---|---|---|---|---|---|---|
| | QP 22 | QP 27 | QP 32 | QP 37 | QP 22 | QP 27 | QP 32 | QP 37 |
| BirdsInCage | 5 | 4 | 4 | 3 | 5 | 5 | 5 | 5 |
| DuckAndLegs | 5 | 5 | 5 | 5 | 5 | 5 | 5 | 5 |
| Kimono | 5 | 5 | 5 | 4 | 5 | 5 | 5 | 4 |
| OldTownCross | 5 | 5 | 4 | 4 | 5 | 5 | 5 | 4 |
| ParkScene | 5 | 4 | 3 | 2 | 5 | 5 | 5 | 4 |
| Traffic | 5 | 5 | 5 | 4 | 5 | 5 | 5 | 5 |

**Rounded Mean Opinion Score (Spatiotemporal Subjective Evaluation) – Pixel-PAQ versus IDSQ**

| Sequence | YCbCr 4:2:2 All Intra | | | | YCbCr 4:2:2 Random Access | | | |
|---|---|---|---|---|---|---|---|---|
| | QP 22 | QP 27 | QP 32 | QP 37 | QP 22 | QP 27 | QP 32 | QP 37 |
| BirdsInCage | 5 | 5 | 4 | 2 | 5 | 5 | 4 | 3 |
| DuckAndLegs | 5 | 5 | 4 | 3 | 5 | 5 | 4 | 3 |
| Kimono | 5 | 5 | 4 | 3 | 5 | 5 | 4 | 2 |
| OldTownCross | 5 | 4 | 3 | 2 | 5 | 5 | 4 | 3 |
| ParkScene | 5 | 5 | 3 | 2 | 5 | 5 | 4 | 2 |
| Traffic | 5 | 5 | 3 | 2 | 5 | 5 | 5 | 4 |

**Rounded Mean Opinion Score (Spatiotemporal Subjective Evaluation) – Pixel-PAQ versus IDSQ**

| Sequence | YCbCr 4:4:4 All Intra | | | | YCbCr 4:4:4 Random Access | | | |
|---|---|---|---|---|---|---|---|---|
| | QP 22 | QP 27 | QP 32 | QP 37 | QP 22 | QP 27 | QP 32 | QP 37 |
| BirdsInCage | 5 | 5 | 4 | 3 | 5 | 5 | 5 | 4 |
| DuckAndLegs | 5 | 5 | 5 | 5 | 5 | 5 | 5 | 5 |
| Kimono | 5 | 5 | 5 | 4 | 5 | 5 | 5 | 4 |
| OldTownCross | 5 | 5 | 4 | 4 | 5 | 5 | 4 | 4 |
| ParkScene | 5 | 5 | 4 | 3 | 5 | 5 | 4 | 4 |
| Traffic | 5 | 5 | 4 | 3 | 5 | 5 | 4 | 4 |

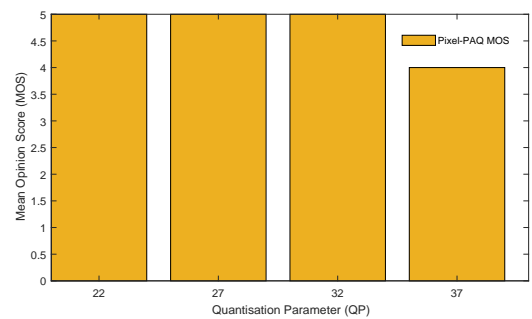
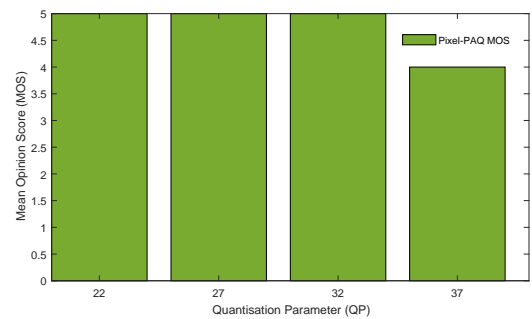

(a)      (b)

**Figure 12.** Two bar graphs which show the Mean Opinion Score (MOS) for the proposed technique over four QP data points (i.e., QPs 22, 27, 32 and 37). Subfigure (a) shows the MOS for Pixel-PAQ versus IDSQ on the BirdsInCage 4:4:4 10-bit sequence using the RA configuration. Subfigure (b) shows the MOS for Pixel-PAQ versus the raw video data on the Traffic 4:4:4 10-bit sequence.

**Table 9.** The MOS results, rounded to the nearest integer, of four participants in the subjective evaluations for Pixel-PAQ versus the raw data. The sequences viewed are coded with initial QPs 22, 27, 32 and 37 on the YCbCr 4:2:0, 4:2:2 4:4:4 versions of the following sequences: BirdsInCage, DuckAndLegs Kimono, OldTownCross, ParkScene and Traffic compressed using the AI and RA configurations. The AI results are shown on the left; the RA results are shown on the right.

**Rounded Mean Opinion Score (Spatiotemporal Subjective Evaluation) – Pixel-PAQ versus Raw Data**

| Sequence | YCbCr 4:2:0 All Intra | | | | YCbCr 4:2:0 Random Access | | | |
|---|---|---|---|---|---|---|---|---|
| | QP 22 | QP 27 | QP 32 | QP 37 | QP 22 | QP 27 | QP 32 | QP 37 |
| BirdsInCage | 4 | 3 | 2 | 2 | 5 | 4 | 4 | 3 |
| DuckAndLegs | 5 | 4 | 3 | 2 | 5 | 5 | 4 | 4 |
| Kimono | 4 | 4 | 3 | 2 | 5 | 4 | 3 | 2 |
| OldTownCross | 5 | 4 | 3 | 2 | 5 | 5 | 4 | 3 |
| ParkScene | 4 | 3 | 2 | 1 | 4 | 3 | 3 | 2 |
| Traffic | 5 | 5 | 4 | 2 | 5 | 5 | 4 | 4 |

**Rounded Mean Opinion Score (Spatiotemporal Subjective Evaluation) – Pixel-PAQ versus Raw Data**

| Sequence | YCbCr 4:2:2 All Intra | | | | YCbCr 4:2:2 Random Access | | | |
|---|---|---|---|---|---|---|---|---|
| | QP 22 | QP 27 | QP 32 | QP 37 | QP 22 | QP 27 | QP 32 | QP 37 |
| BirdsInCage | 4 | 3 | 2 | 1 | 5 | 4 | 3 | 2 |
| DuckAndLegs | 5 | 5 | 4 | 3 | 5 | 5 | 4 | 3 |
| Kimono | 5 | 4 | 2 | 1 | 5 | 4 | 3 | 2 |
| OldTownCross | 5 | 4 | 2 | 1 | 5 | 4 | 2 | 2 |
| ParkScene | 5 | 4 | 2 | 1 | 5 | 5 | 4 | 2 |
| Traffic | 5 | 4 | 3 | 2 | 5 | 5 | 4 | 2 |

**Rounded Mean Opinion Score (Spatiotemporal Subjective Evaluation) – Pixel-PAQ versus Raw Data**

| Sequence | YCbCr 4:4:4 All Intra | | | | YCbCr 4:4:4 Random Access | | | |
|---|---|---|---|---|---|---|---|---|
| | QP 22 | QP 27 | QP 32 | QP 37 | QP 22 | QP 27 | QP 32 | QP 37 |
| BirdsInCage | 4 | 3 | 2 | 1 | 5 | 4 | 3 | 2 |
| DuckAndLegs | 5 | 5 | 4 | 3 | 5 | 5 | 4 | 3 |
| Kimono | 5 | 4 | 3 | 2 | 5 | 5 | 4 | 3 |
| OldTownCross | 5 | 4 | 2 | 1 | 5 | 5 | 4 | 3 |
| ParkScene | 5 | 4 | 3 | 2 | 5 | 4 | 4 | 3 |
| Traffic | 5 | 5 | 4 | 2 | 5 | 5 | 4 | 4 |

The MOS results tabulated in Table 8 indicate that there are negligible perceptual differences between the Pixel-PAQ coded versions versus the IDSQ coded versions of the following YCbCr 4:2:0 (8-bit), 4:2:2 (10-bit) and 4:4:4 (10-bit) sequences: BirdsInCage, DuckAndLegs, Kimono, OldTownCross, ParkScene and Traffic. This consistently proved to be the case for the AI and RA QP = 22 and QP = 27 tests. In most of the Pixel-PAQ versus IDSQ AI QP = 37 tests, including the experiments undertaken on the BirdsInCage 4:4:4 and 4:2:2 sequences, the chrominance quantisation-induced distortions are noticeable in the versions coded by Pixel-PAQ. Due to the fact that BirdsInCage comprises a high level of saturation and also a combination of low variance and high variance Cb and Cr data, the coarser quantisation applied to the chroma Cb and Cr CBs by Pixel-PAQ is conspicuous in the AI QP = 37 tests. The most significant bitrate reduction achieved by Pixel-PAQ, in comparison with IDSQ, is attained on the 4:4:4 version of the BirdsInCage sequence (75% bitrate reduction in the RA QP = 22 test). Furthermore, visually lossless coding is accomplished in addition to achieving the vast bitrate reduction.

The data recorded in Table 9 tabulates the subjective evaluation MOS scores for the Pixel-PAQ coded sequences versus the raw video data. The goal in this set of subjective evaluations is to ascertain if Pixel-PAQ successfully achieves visually lossless coding. The subjective evaluation participants confirm that visually lossless coding is accomplished by Pixel-PAQ in the AI QP = 22 and RA QP = 22 tests in the vast majority of cases and, in some cases, in the AI QP = 27 and the RA QP = 27 tests. Irrespective of the SSIM and PSNR values tabulated in Table 4 to Table 7, the perceptual quality of the sequences coded by Pixel-PAQ using the GOP-based (inter coding) Random Access encoding configuration is considerably superior to those coded using the All Intra (intra-only) encoding configuration. This is especially true according to the tests performed with the higher initial QP values (i.e., QPs 32 and 37). The main reason for this is because intra-only coding is a relatively primitive spatial image coding method that does not employ motion estimation, motion compensation, advanced motion vector prediction and bidirectional inter prediction. Coding with I-frames only does not take into account the temporal redundancies that exist between frames, which constitutes a major shortcoming of All Intra lossy coding. Conversely, motion data with GOP-based inter coding in HEVC can be signalled to the decoder with the utilisation of merge mode or by motion vector differences, picture reference indices and the direction of the inter prediction. With this in mind, in the sequences coded by Pixel-PAQ, the quantisation-induced compression artifacts proved to be vastly more conspicuous in all AI tests, especially so at initial QP = 37. As such, based on the subjective evaluation results, it can be inferred that GOP-based inter coding is considerably more effective than intra-only coding in JND-based lossy video coding applications (and also lossy video coding applications in general).

**4.0 Conclusions**

A novel CB-level, JND-based luma and chroma perceptual quantisation technique is proposed for HEVC (named Pixel-PAQ). Pixel-PAQ exploits HVS-based perceptual masking, whereby spatial CSF-based luminance masking and chrominance masking are employed to achieve JND-based perceptual quantisation and visually lossless coding. The QPs for the Y CB, the Cb CB and the Cr CB are perceptually increased in order to considerably reduce bitrates without incurring a conspicuous impact on the reconstruction quality of the compressed video data. In the subjective and objective evaluations, Pixel-PAQ is compared with a state-of-the-art JND-based perceptual quantisation technique based on luminance masking (named IDSQ). In comparison with IDSQ, Pixel-PAQ achieves vast bitrate reductions — of up to 75% (68.6% over four QP data points) — on the YCbCr 4:4:4 10-bit version of the BirdsInCage sequence. In addition to this, the subjective evaluations confirmed that visually lossless coding is achieved in almost all cases in which the initial QP = 22 (for both the AI and RA tests). This proved to be the case for the Pixel-PAQ versus IDSQ tests and also the Pixel-PAQ versus raw video data tests. Finally, no significant differences in encoding and decoding runtimes are observed for Pixel-PAQ versus IDSQ; Pixel-PAQ achieved marginal runtime reductions in all tests.

**Acknowledgements**

The author would like to sincerely thank Shevach Riabtsev — Senior Video Engineer at Beamr — for his valuable feedback. With Shevach's extensive knowledge and experience in terms of working on state-of-the-art HEVC algorithms in addition to his contributions accepted by JCT-VC during the standardisation of HEVC, the valuable comments provided by Shevach are greatly appreciated by the author.

The author would also like to show gratitude to Victor Sanchez, who is presently acting as the author's PhD supervisor in the Department of Computer Science at the University of Warwick. As first author, Victor has published several papers in high impact IEEE journals and major IEEE conferences on the topic of mathematically lossless intra prediction coding in HEVC.